\documentclass[pre,showpacs,preprintnumbers,onecolumn]{revtex4}
\usepackage{amssymb}
\usepackage{color}
\usepackage{graphicx}
\usepackage{subfigure}
\usepackage{multirow}
\usepackage{enumerate}

\begin{document}

\title{Design principles of noise-induced oscillation in biochemical reaction networks: II. coupled positive and negative feedback loops} 
\author{Jaewook Joo and Sanjeev Chauhan}
\affiliation{Department of Physics and Astronomy, University of Tennessee, Knoxville TN}

\begin{abstract}
According to the chemical reaction network theory, the topology of a certain class of chemical reaction networks, regardless of the kinetic details, sets a limit on the dynamical properties that a particular network can potentially admit; the structure of a network predetermines the dynamic capacity of the network. We note that stochastic fluctuations can possibly confer a new dynamical capability to a network. Thus, it is of tremendous value to understand and be able to control the landscape of stochastic dynamical behaviors of a biochemical reaction network as a function of network architecture. Here we investigate such a case where stochastic fluctuations can give rise to the new capability of noise-induced oscillation in a subset of biochemical reaction networks, the networks with only three biochemical species whose reactions are governed by mass action kinetics and with the coupling of positive and negative feedback loops. We model the networks with master equations and approximate them by using linear noise approximation. For each network, we read the signal-to-noise ratio value, an indicator of amplified and coherent noise-induced oscillation, from the analytically derived power spectra. We classify the networks into three performance groups based on the average values of signal-to-noise ratio and robustness. We identify the common network architecture among the networks belonging to the same performance group, from which we learn that the coupling of negative and positive feedback loops generally enhance the noise-induced oscillation performance better than the negative feedback loops alone. The performance of networks also depends on the relative size of the positive and negative feedback loops; the networks with the bigger positive and smaller negative feedbacks are much worse oscillators than the networks with only negative feedback loops. 

\end{abstract}

\maketitle

\section{Introduction}

In molecular systems biology, the biochemical reaction networks represent the complex biological systems with a large number of biological components and many interactions among themselves. The examples of the networks are gene regulatory networks, signal transduction networks, and metabolic networks which process the cellular information, make the cell-fate determining decision, and are inherently coupled together. These networks are subject to noise from various sources, i.e., intrinsic and extrinsic noise. The dynamical behaviors of those networks determine the physiology and phenotype of the living organimsms. It is of great importance that we deepen our understanding of the interplay between the noise, the structual properties of the biochemical reaction network, and the resulting dynamics. This new knowlege enables us to understand the mechanims of how the signaling pathways work in natural organisms, to intervene those pathways to modify the dynamics of genes and proteins, and to design a new synthetic circuit with a desired functionality. Our current paper is to pursue the deeper understanding of the design principles of stochastic oscillations arising from biochemical reaction networks. Presently our work is limited to small-sized networks, but lays a good foundation for further generalization to an arbitrarily-sized network. 

The interplay between the structural properties of the chemical reaction networks and the dynamics that the networks can potentially admit has been deeply studied in the chemical reaction network theory (CRNT)~\cite{Feinberg1979,Gunawardena2003}. In the deficiency-zero theorem of the CRNT, any weakly reversible network with both mass action kinetics and zero dificiency is proven to have only one positive and locally asymptotically stable statedy state~\cite{Feinberg1987}. The structral properties of a chemical reaction network specified by the reversability and the deficiency of the network, regardless of the kinetic details, sets a limit on the qualitative dynamical properties that that particular network can potentially admit. However, because the deficiency theorems are constructed on the strict conditions, their usefulness to the systems and network biology are limited. Moreover, it has not been explored how the structure of the chemical reaction networks influence the dynamical capacities of the stochastic chemical reaction networks. Stochastic fluctuations are ubiquitous and prevalent in a cellular environment and consequently influence the operation and functioning of the gene regulatory and cell signaling networks. Those fluctuations can possibly confer a new dynamical capability to the network or destroy an existing dynamical capacity: e.g., noise-induced bistability, noise-indued stabilization, noise-induced synchronization, and noise-induced oscillation~\cite{Samoilov2005, Bishop2009, Qian2011,Turcotte2008, Parker2011, Athreya2012, Kurebayashi2012, Medvedev2010}. Thus, it is of tremendous value to systematically define the landscapes of stochastic dynamical behavior as a function of the structural properties of the biochemical reaction networks. 

Among many cellular dynamical phenomena, this paper concerns the biochemical oscillation. Oscillations are prevalent in cellular biology partly because of the living organisms' adaptation or readiness to a periodic or abrupt environmental change. The few examples of the cellular biochemical oscillations are cell cycle, circadian rhythm, NF-$\kappa$B, p53, developmental clock, neural rhythms, and hormone~\cite{Tyson, Goldbeter}. To quantitatively understand such periodic phenomena, many theoretical models have been put forth and extensively studied. The previous studies revealed a few important requirements for biochemical oscillators~\cite{Novak}: i) negative feedback loops with sufficient time delay, ii) non-linearity of kinetic laws, and iii) appropriate balancing of synthesis and degradation rates of chemical species. One of the above three restrictions is the structral condition of the underlying networks: negative feedback loops with time delay. A negative feedback loop is the cyclic pathway consisting of the odd number of inhibitory edges. The time delay can be realized in the networks in two different manners: One by having the explicit time delay in biochemical interaction and another by having positive feedback loops in addition to the negative feedback loops. Our primary questions in this paper are how the noise relaxes or tightens the above three conditions for biochemical oscillations. Particularly, we are very keen to {\it enact the new requirements} for stochastic biochemical oscillators and to compare the stochastic oscillatory behaviors between networks with negative feedback loops alone and the networks with coupled positive and negative feedback loops. 

In our previous work~\cite{Joo1}, we investigated the requirements for stochastic biochemical oscillators and compared the networks with only negative feedback loops. The networks consist of three biochemical species governed by mass action kinetics and are allowed to have only negative feedback loops. We proved that the negative feedback loops are required for stochastic oscillation in these small-sized networks with mass action kinetics. Then, we numerically demonstrated that all the networks have one positive and locally stable steady state and the stochastic fluctuations enable all of them to exhibit prominent stochastic oscillations, i.e., coherence resonance or noise-induced oscillation, in the various biologically feasible parameter ranges~\cite{Gang, Wiesenfeld, Pikovsky, Kuske, Mckane1, Mckane2}. Stochastic fluctuations confer an additional dynamical capacity of stochastic oscillation to the group of biochemical reaction networks with negative feedback loops.

Numerous biochemical oscillators are equipped with coupled positive and negative feedback loops~\cite{Tsai, Tian, KimJ, KimD}. The few examples are mitiotic trigger in Xenopus~\cite{Pomerening}, $Ca^{2+}$ spikes/oscillations~\cite{Keizer}, circadian clock~\cite{Paranjpe}, galactose-signaling network in yeast~\cite{Acar}, and p53-Mdm2 oscillator~\cite{Wee}. The addition of positive feedback loops to the biochemical oscillators can confer some functional and performance advantages ~\cite{Tsai, Tian, KimD, KimJ}. The networks with the positive-negative interlinked feedbacks can have the inreased frequency-tunable range and the enhanced robustness of biochemical oscillations~\cite{Tsai}. Additionally, the positive-negative interlinked feedback loops can give rise to a variety of dynamical behaviors such as monostability, bistability, exitability, and oscillations in the change of relative strength of the feedback loops~\cite{Tian} and even make the cellular signal responses more desirable in noise environment~\cite{KimD}. It seems that the nature must have forced the living organisms to evolve to favor the networks with interlinked postive and negative feedback loops because of their many functional advantages.

In this present paper, we add one more favorable point to the long list of the advantage of networks with positive and negative interlinked feedbacks: networks with positive-negative coupled feedbacks are much better noise-induced oscillators than networks with only negative feedback loops in various biologically feasible parameter ranges. Again, we consider small-sized networks with mass action kinetics and two types of repression: repression by proteolysis and repression by transcriptional control. We generate the exhaustive list of all possible network structures with interlnked feedback loops, resulting in the sixty-three networks with the differential coupling of positive and negative feedback loops, and model each network with a chemical master equation which is approximated to a linear Fokker-Planck equation through van Kampen's system size expansion. Firstly, stochastic fluctuations indeed confer a new dynamical capacity of stochastic oscillation to all the networks with interlinked feedback loops. Secondly, implementing a K-medoids clustering algorithm, we group the sixty-three networks into three performance groups based on the average values of signal-to-noise ratio and robustness and identify the common network architecture among the networks belonging to the same performance group. We learn that the coupling of negative and positive feedback loops (PNFBL) generally enhance the noise-induced oscillation performance better than the negative feedback loops (NFBL) alone. However, the performance of PNFBL networks depends on the size of the positive feedback loops (PFBL) relative to that of the NFBL in the networks; the performance of the networks with the bigger PFBL than NFBL is worse than that of the networks with only NFBL. We single out two networks which stand out in their performance of noise-induced oscillation in all four different parameter ranges and for two different models of repression. Thirdly, we also elucidate the machanisms of noise-induced oscillation and find that for most networks, the natural frequency sets the low bound of the resonant frequency of the noise-induced oscillation. This corroborates the present understanding of the noise-induced oscillation mechanisms; the imaginary part of a complex eigenvalue from the Jacobian matrix of a linearized network is responsible to generate an inherent rotation in the network and the noise enhances such a rotation. However, we also find that a few networks with purely real eigenvalues can admit a very prominent and amplified noise-induced oscillation.

The paper is organized in the following way. In the result section, we discuss the K-medoids clustering of sixty-three networks into three performance groups, the classification of networks by network structure, the identification of a common network structural signature per performance group, the distributions of signal-to-noise ratio from real and complex eigenvalues of Jacobian matrices, and finally the correlation between the values of signal-to-noise ratio and the spiralness and proximity. In model and method section introduced all the networks under our consideration, the derivation of the power spectrum from the chemical master equation based on the van Kampen system size expansion, parameter sampling method, definition of signal-to-noise ratio, prominence and robustness, the eigenvalue calculation, and the linear stability of the networks.

\section{Results}

\begin{figure}[h!]
\begin{center}
\includegraphics[height=10cm, width=15cm]{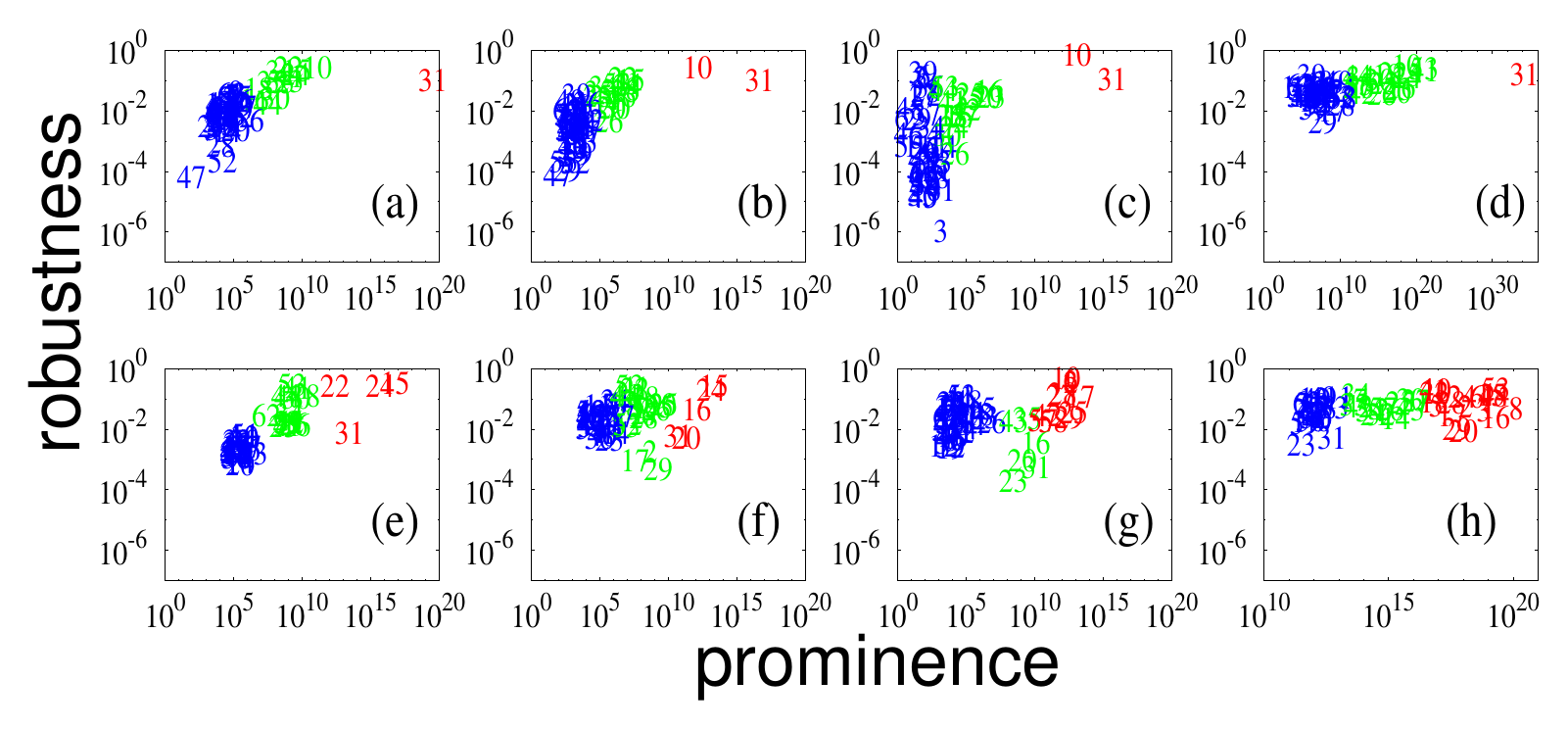}
\end{center}
\caption{Classification of sixty-three biochemical reaction networks into three performance groups, based on the values of prominence and robustness and using K-medrid clustering algorithm. In top row (a)-(d), the repression is modeled by proteolysis. The proteolysis rate is sampled from the following four intervals: a) $(10^{-7},10^{-5})$, b) $(10^{-5},10^{-3})$, c) $(10^{-3},10^{-1})$ and d) $(10^{-7},10^{-1})$. In bottom row, repression is modeled by transcriptional control. The transcriptional repression rate is sampled from the following four intervals: e) $(10^{-3},10^{-1})$, f) $(10^{-1},10)$, g) $(10,10^3)$, and h) ($10^{-3},10^3)$. The values of all the other kinetic rate constants are sampled from a preset respective biologically feasible interval for a)-c) and e)-g) whereas all of them are sampled from the same interval for d) and h).} 
\label{Fig1}
\end{figure}

The exhaustive enumeration of all the networks with three nodes and with at least one negative feedback loop reveals that there are sixty-three biochemical reaction networks. Each of sixty-three networks has one positive and locally stable steady state in the chosen parameter ranges. If a network admits an unstable steady state, then that set of parameter values is resampled until the network admits the stable fixed point. The statics of resampling cases is provided in the supplimentary information. Sixty-three biochemical reaction networks are classified, based on their values of prominence and robustness. Prominence is a measure of the coherence and amplification of a stochastic oscillation, defined as the averaged maximum signal-to-noise ratio (SNR). Robustness is the fraction of the sample points that admit the value of the maximum SNR greater than one. The maximum SNR means that the largest SNR value among three chemical species. See the method section for the detailed definition of prominence and robustnss and how to calculate them. To classify the networks, we use a machine-learning classification tool, K-medoids clustering. The kinetic rate constant values are sampled in two different manner, biologically and non-biologically. For biologically feasible parameter values, all the kinetic rate constants are sampled from a preset biologically feasible range in Table IV. Particularly, we consider three different intervals of the repression strength corresponding to weak, intermediate, and strong repression. We do not know even the approximate values of repression strength and want to see the dependence of noise-induced oscillation performance on repression strength. For non-biological parameter values, all the kinetic rate constants are sampled from the same "non-biological" range. This is to see the overall behavior of the networks across a parameter space.

\begin{figure}[h!]
\begin{center}
\includegraphics[height=10cm,width=15cm]{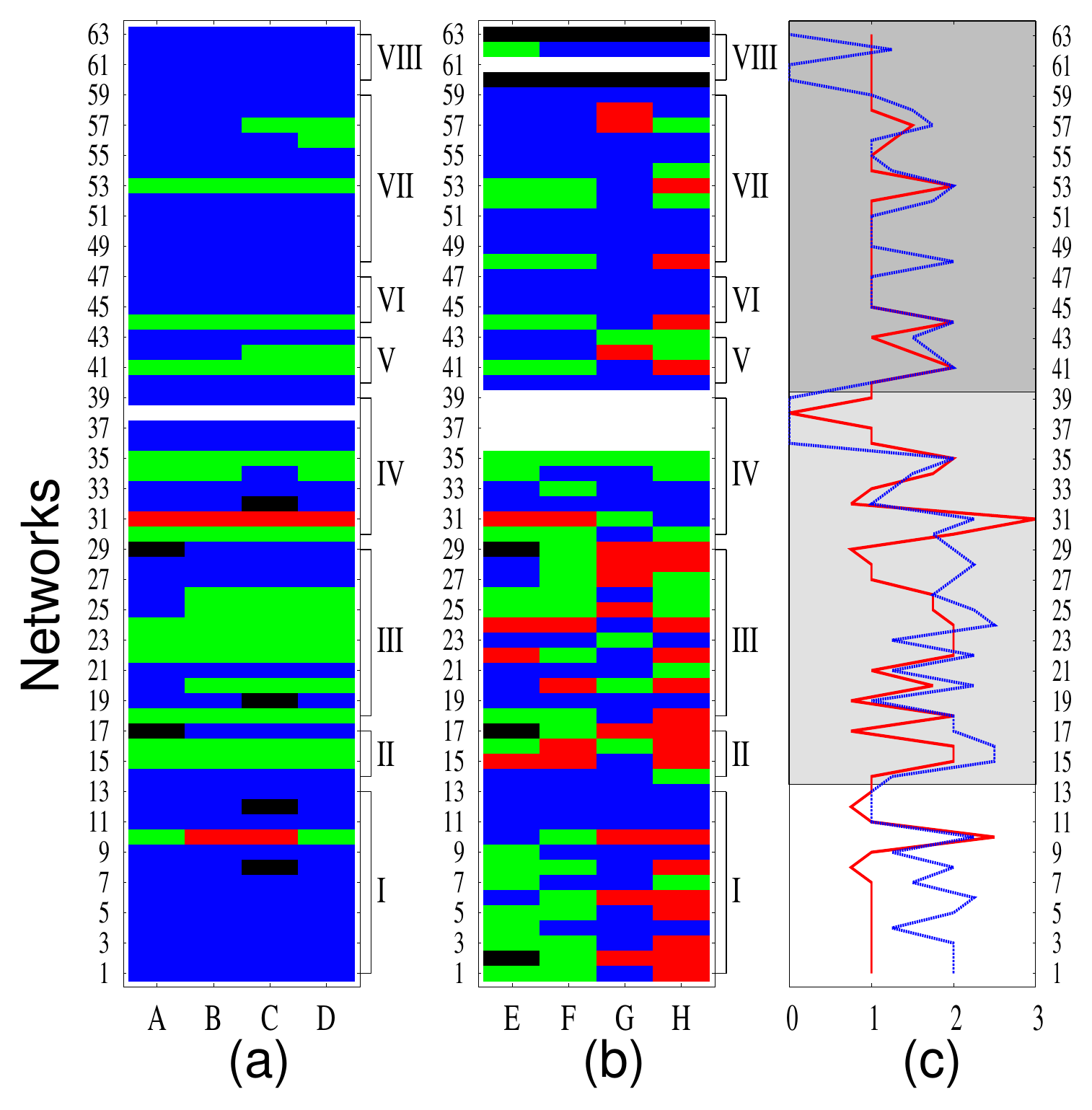}
\end{center}
\caption{Relationship bewteen stochastic oscillatory performance and network structure. The sixty-three networks are classified according to network structural properties, i.e., the number and the size of negative and positive feedback loops: architectural classes I through VIII. For each network, we consider four different sampling intervals and two repression models. In the left map, repression is modeled by proteolysis whereas in the middle map, repression is by transcriptional control. The four differerent sampling intervals are labeled as A through D for repression by proteolysis and as E through H for repression by transcriptional control. Vertical axis shows the classification of networks according to network architecture as that given in Tables \ref{models} and \ref{models2}. While color indicates the absence of data due to technical issues. Black color denotes that all sample points yield $SNR\le 1$. Red, green or blue color is assigned to each network in each parameter sampling range and exactly corresponds to the classification from the $K$-medoids clustering in Figure 1. In the right graph is the average performance value for each network: integer numbers, 0,1,2,3 are assigned to four colors, black, blue, green, and red, respectively. Red line indicates the average performance value from (a) repression by proteolysis whereas blue line denotes the performace value from (b) repression by transcriptional control.}
\label{Fig2}
\end{figure}

\begin{figure}[h!]
\begin{center}
\includegraphics[height=10cm,width=15cm]{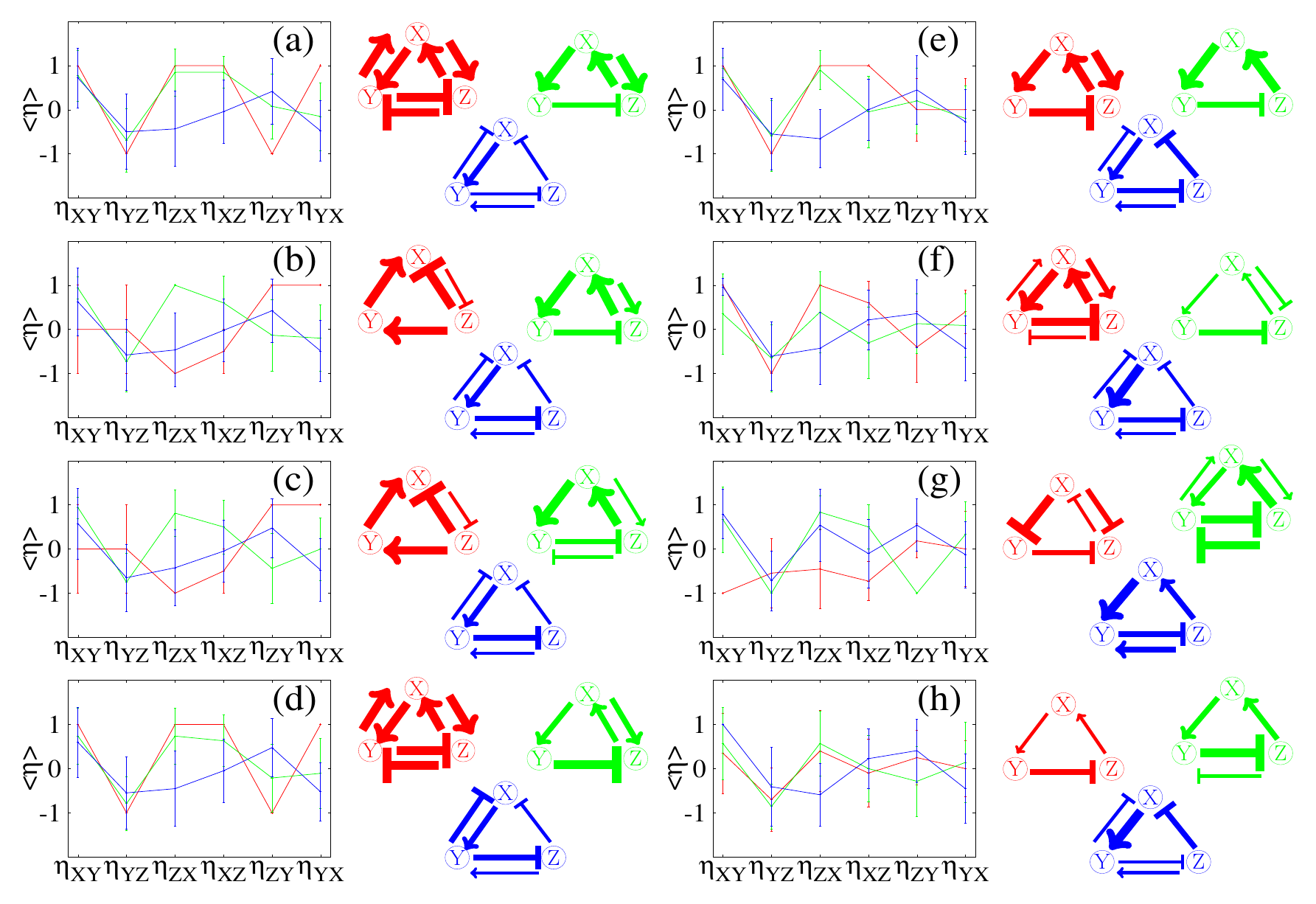}
\end{center}
\caption{Identification of common network architectural properties among the networks in the same performance group. The subfigures (a)-(d) in the left two columns are for repression by proteolysis and the subfigures (e)-(h) in the right two columns for repression by transcriptional control. $\eta_{ij}$, where $i,j=X,Y,Z$, represents the numerical value of a directed edge going from a node $i$ to a node $j$ and $\eta_{ij} \in \{-1,1\}$. E.g., $\eta_{XY}=-1$ indicates X represses Y while $\eta_{XY}=+$ denotes X activates Y. $\langle \eta \rangle$ denotes the average value of $\eta_{ij}$ over the properly rotated networks belonging to the same performance group. The vertical bars indicate the standard deviations. The average values of $\langle \eta \rangle$ are visualized in the third and fourth columns. Networks drawn adjacent to the boxes are the pictorial representations of common architecture of the networks in the same performance group. The same color is used for the common networks as that used in Figure \ref{Fig1}. Type and thickness of the lines are determined by the average value of $\eta_{ij}$ as discussed in the methods section. Its positivity (negativity) denotes that the type of line is activation (repression).}
\label{Fig3}
\end{figure}

Fig.~\ref{Fig1} shows the classification of the sixty-three networks into three performance groups. The performance of networks is based on the values of prominece and robustness of stochastic oscillation.  We classify the biochemical reaction networks into three clusters, using $K$-medoids clustering. The K-medroid performs robustly and the choice of K=3 is well justified by the occurrence of a dramatic drop at K=3 in the graph of sum distance (error) versus the numebr of clusters over almost all parameter ranges. The graph of sum distance versus the number of clusters for Fig.~\ref{Fig1} is provided in the Fig.~\ref{Fig8} in the supplimentary information. Just like other searching algorithms, the K-metroid clustering algorithm can be trapped in local mimina. To prevent such a trap, we start from many different initial conditions and optimize the performance of the K-metriod clustering algorithm. Because of techanical issues, we could not evaluate one network (network 38) for the case of repression by proteolysis and five networks (networks 36-39, 61) for the case of repression by transcriptional control. We label three clusters of networks as follows: red as the best performing networks, green as the medium performers, and blue as the worst performers. In Fig.~\ref{Fig2} and \ref{Fig3}, we will indentify the common network structural properties among the networks belonging the same color group. The clustering also enables us to identify the all-season best performer of stochastic oscillation, network 31. This network 31 belong to the red group for both repression models and across four different sampling intervals of repression strength. The Network 31 consists of two three-dimensional NFBL plus  three two-dimensional PFBL. One interesting observation is that the clustering is mainly done by the prominence values and does not strongly depend on the robustness. We do notice the decreasing trend of the prominence of most of the networks as the repression strength increases for both repression models in Fig.~\ref{Fig1}. For example, As the repression strength increases from a) to c), the prominence value of network 31 decreases respectively. This pattern is most conspicuously noticed with the networks in green performance group.

In Fig.~\ref{Fig2}, the sixty-three networks are classified into eight different architectural classes by network structural properties as provided in the Tables I and II. This newtork structural classification is based on the number and the size of positive and negative feedback loops. Those eight architectural classes of the sixty-three networks can form three architectural groups in a coarse-grained manner. The architectural group I includes all the networks with only negative feedback loops, Networks 1-13, whereas the architectural group II (classes II through IV) contains the networks with the coupling of smaller postive feedback loops and larger negative feedback loops, Networks 14-39. Finally the architectural group III (classes V through VIII) consists of the networks with the coupling of larger positive feedback loops and smaller negative feedback loops and the networks with linear chains, Networks 40-63. The networks in the architectural group I dominantly belong to blue performance group for the case of repression by proteolysis whereas for repression by transcription control all three colors are mixed. A majority of the networks in the architectural group II belong to green performance group for repression by proteolysis whereas for transcriptional repression, they belong to either green or red performance group with exception of a very few blue performance networks. The networks in the architectural group III, for both cases of repression by proteolysis and transcription control, mostly belong to blue or black performance group. As far as the case of repression by proteolysis is concerned, the blue performance arises from two architectural groups (or four architectural classes): group I (class I) and group III (classes V through VIII). The clear message is that the networks either with exclusively negative feedback loops or with the larger-sized positive feedback loops and the smaller-sized negative feedback loops can admit the noise-induced oscillation in the chosen biologically feasible parameter ranges, but their oscillations are neither sufficiently well amplified nor coherent. In other words, the networks with the coupling of smaller-sized positive feedback loops and larger-sized negative feedback loops can admit quite well amplified and coherent stochastic oscillations. Thus, the recommended network strucure for biochemical oscillators is one of the networks belonging to the architectural group II. 

To see the clear relationship among the individual network structures, the parameter sampling range, and the performance of networks altogether, we represent each network not with the absolute value of average max SNR but with the performance group to which it belongs. Then, we average the performance colors of each network over the four different parameter sampling ranges for two different repression models, as presented in the rightmost subfigure of Fig.~\ref{Fig2}. Since the average performance scores are quite noisy, it is very hard to appreciate any correlation between the individual networks and their average performance scores. But, it is easy to see that the average performance scores and the network architectural groups (represented with three different background darknesses in Fig.~\ref{Fig2}(c) are closely correlated for both repression models (blue and red lines in Fig.~\ref{Fig2}(c)). In other words, the networks belonging to their respective architectural group demonstrate the similar performance, regardless of the detailed repression models.

In Fig.~\ref{Fig3}, we identify the common architectural properties among the networks that are classified into the same performance group as discussed in Figure 1. Each of the networks belonging to the same performance group are rotated until they are all aligned such that the Hamming distance among the networks are minimized with respect to a reference network. We then calculate the average edge value, defined as $\langle \eta_{ij} \rangle$ where $ij$ denotes the directed edge from a node X to a node Y as presented in Fig.~\ref{Fig2}. We graphically represent the average edge values into the common network architecture for each of the combination of two repression models and four parameter ranges. The average edge values are converted to four different thickness of an edge in the graphical representation: $[1,0.75)$ (maximum thickness), $[0.75,0.5)$ (medium thickness), $[0.5,0.25)$ (minimum thickness) and $[0.25,0.0]$ (no edge). On the one hand, the networks belonging to the blue performance group share the following structural properties: one three-dimensional positive feedback loop coupled with many two-dimensional negative feedback loops.  In other words, The blue networks are characerized with the strong presence of multiple two-dimensional (small) negative feedback loops and three-dimensional (large) positive feedback loop. We conclude that the networks with a larger positive feedback loop coupled with small negative feedback loops are likely to belong to the blue (worst) performance group. On the other hand, the networks belonging to the green performance group are characterized by the strong presence of three-dimensional (large) negative feedback loop and one two-dimensional (small) positive feedback loop. In the networks belonging to the Red performance group, we see the same strong presence of three-dimensional negative feedback loop as wellas the stronger presence of multiple two-dimensional positive feedback loops. The best network topology for highly amplified and coherent stochastic oscillators is a large negative feedback loop coupled with many small positive feedback loops. We demonstrate the existence of the network signature for each performance group consistently across two different repression models and over four different intervals of the model parameters. This is a remarkable numerical evidence of the relationship between network architecture and network performance.

\begin{figure}[h!]
\begin{center}
\includegraphics[angle=0, height=10cm,width=15cm]{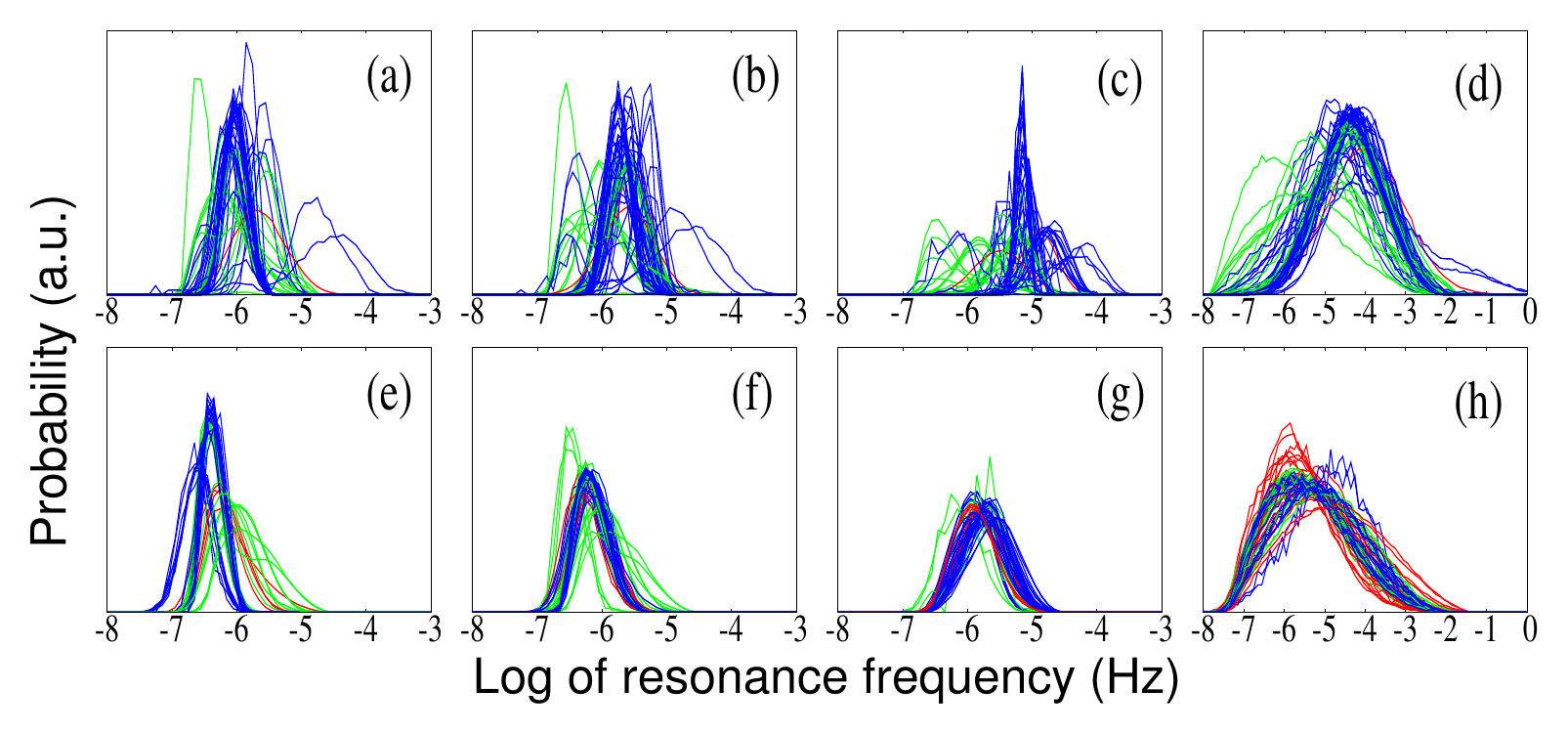}
\end{center}
\caption{Distribution of resonant frequencies of stochastic oscillations from sixty-three networks over two repression models and four parameter sampling ranges.Each color line corresponds to the histogram of resonant frequencies from a network belonging to the same color peformance group defined in Figure 1. The histogram was constructed with $10^6$ parameter samples. The labels (a) through (h) are the same as in Figure 1.}
\label{Fig4}
\end{figure}

In Fig.~\ref{Fig4}, we present the distributions of the resonant frequencies of stochastic oscillations from the sixty-three networks which are colored according to their performance group as defined in Fig.~\ref{Fig1}. The resonant frequency is defined as a non-zero frequency at which the power spectrum peaks. The distribution is plotted with a subset of power spectra which has a peak at non-zero frequency, i.e., a fraction of $10^6$ sample points that yield maximum SNR$\ge$ 1. Since any noise-free network has a single stable fixed point, the existence of a resonance frequency is purely due to noise-induced effect. For the case of repression by transcriptional control (e)-(h), the distributions of resonant frequencies are substantially overlapped over almost all of the networks, independent of their performance and network architectures. For the case of repression by proteolysis (a)-(d), the distributions are not as homogeneous as for the case of repression by transcriptional control. The blue-colored networks tend to have more homogeneous distribution than the green-colored networks. For both repression models, the peaks of resonant frequency distributions are shifted to the right as the repression strength increases.  This is a clear signal that the noise-induced resonant frequency is positively correlated with the repression strength. The range of resonant frequencies falls within the biologically relevant range, between hours (corresponding to $10^{-4} Hz$) to years (corresponding to $10^{-8} Hz$).

\begin{figure}[h!]
\begin{center}
\includegraphics[angle=0, height=10cm,width=15cm]{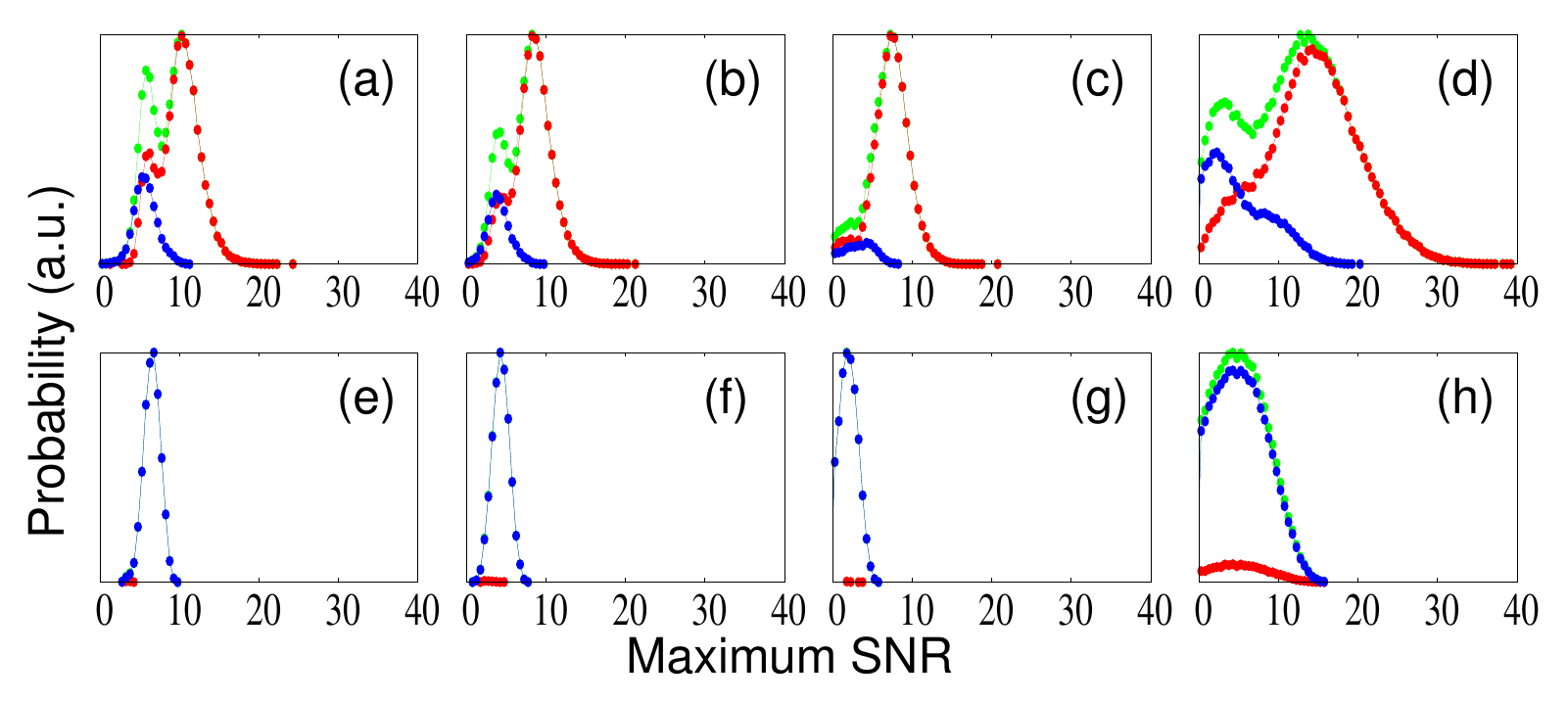}
\end{center}
\caption{Distribution of the Logarithm of the maximum signal-to-noise ratio values for network 31 and network 44 with repression by proteolysis. On x-axis is the Log of maximum SNR values whereas on y-axis is the frequency of the maximum SNR values. The blue cuvres denote the distribution of the maximum SNR contributed from the purely real eigenvalues of the jacobian matrix. The red curves depicks the distribution from the complex eigenvalues. Finally, the green curves are the total sum of both distributions from both real and complex eigenvalues.}
\label{Fig5}
\end{figure}

\begin{figure}[h!]
\begin{center}
\includegraphics[height=5cm,width=15cm]{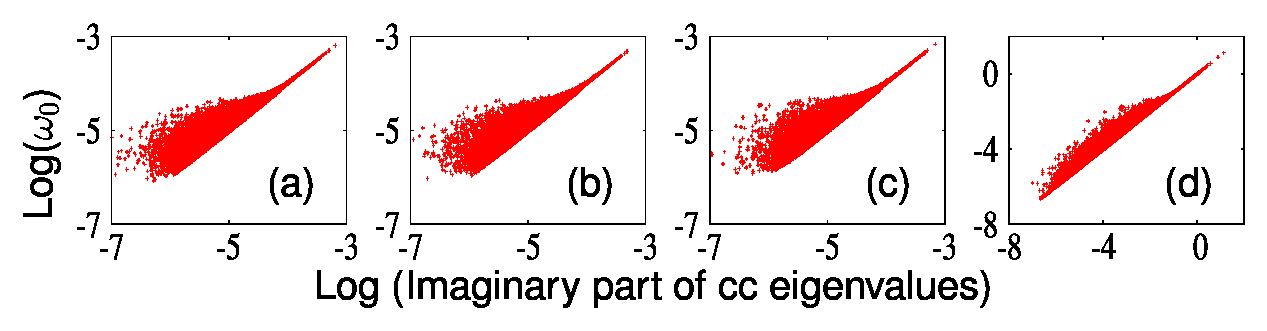}
\includegraphics[height=5cm,width=15cm]{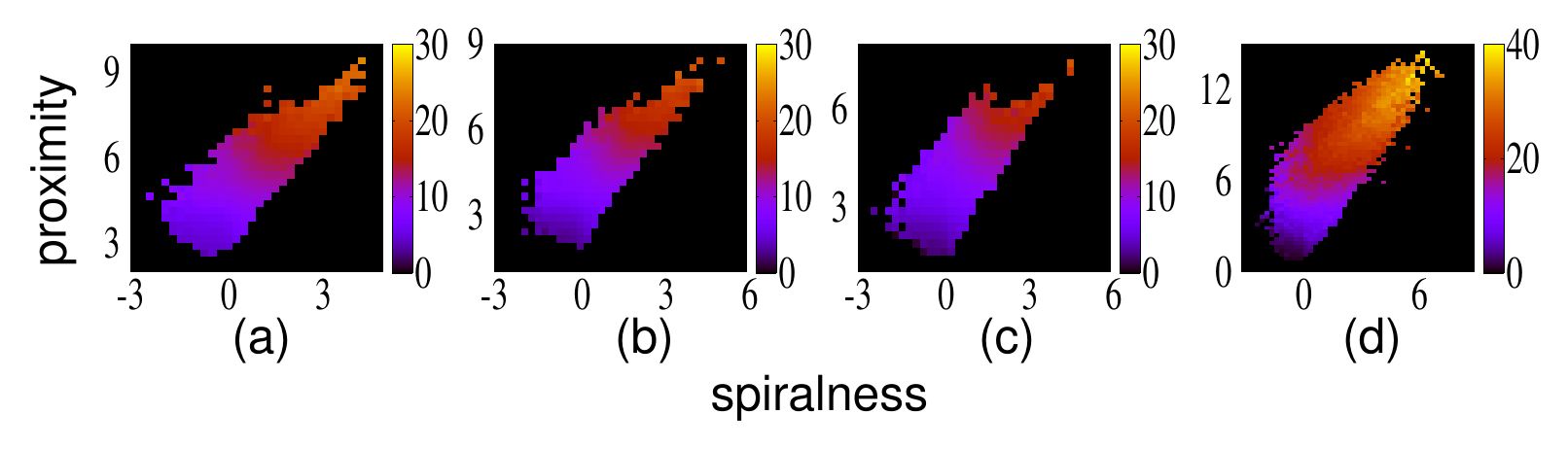}
\end{center}
\caption{Mechanisms of noise-induced oscillation that arises from the complex eigenvalues of the Jacobian matrix. In the top panel is depicted the correlation between the imaginary part of the complex eigenvalues of the Jacobian matrix and the resonant frequency of stochastic oscillation for the case of network 31 with repression by proteolysis. Each subfigure contains about $10^6$ dots which represent the sets of randomly sampled parameter values. In the bottom panel is provided the plot of spiralness versus proximity for the same Network 31 with repression by proteolysis. Both axes are is in logarithmic scale. The colored heat map indicates the logarithmic value of maximum SNR.} 
\label{Fig6}
\end{figure}

In Fig.~\ref{Fig5} are discussed the origins and mechanisms of the stochastic oscillations accompanied with the large maximum SNR values. Calculating the discriminant of a Jacobian matrix with a set of randomly sampled parameter values for a chosen network, we accurately determine if all three eigenvalues of the 3 x 3 Jacobian matrix are real or the mixture of real and complex conjugate pairs. We repeat the discrimimant calculation with the $10^6$ sets of randomly sampled parameter values from four different parameter ranges for the sixty-three networks with two different repression models. Each subfigure in Fig.~\ref{Fig5} presents three histograms of the maximum SNR values: one from the samples yielding all real eigenvalues, another from the samples resulting in complex conjugate pairs, and the last from the total samples. The distributions from only networks 31 and 44 are presentd in the main text, but the distributions from the rest of the networks are provided in the Figs.~\ref{Fig9} and \ref{Fig10} in the supplimentary information. Note that the log of the maximum SNR values less than 0 do exist, but they are not presented here. For the case of network 31 and almost all of the networks in the network architectural group II (networks 14-39), the complex eigenvalues make a dominant contribution to the distribution of maximum SNR values. Particularly the high values of maximum SNR come exclusively from the samples yielding the complex eigenvalues. The imaginary part of the complex eigenvalues is indicative of a rotational flow in a deterministic system and the noise tends to amplify that rotational motion, resulting in the maximally amplified and coherent oscillation. So, this high correlation between the existence of complex eigenvalues and the high values of maximum SNR is well understood. The further discussion of the complex eigenvalue cases will follow in Fig.~\ref{Fig6}. A significant contribution of the real eigenvalues to the maxum SNR values is detected from network 44. All the networks belonging to the network architectural group III (networks 40-63) benefit the similar huge contribution from the real eigenvalues as shown in Fig.~\ref{Fig9} in the supplimentary information. It is worthwhile to note that only a very few networks belonging to this third architectural group are in green performance group, but a majority of the networks are in blue performance group. This seems to suggest that the real eigenvalues do not elicit as large SNR values as the complex eigenvalues do. However, the network 44 belongs to the green performance group and its intermediate performance is exclusively due to purely the real eigenvalues, leaving us puzzled. Finally, the networks from the first network architectural group I (networks 1-13) have the very similar distributions from both real and complex eigenvalues. The above obervations are primarily based on the proteolysis repression model as shown in Fig.~\ref{Fig9} in the supplimenary information. For the case of repression by transcriptional control, all of sixty-three networks dominantly produce the complex eigenvalues across four different parameter sampling ranges. The network architecture seems to determine the ratio between twos distributions from real and complex eigenvalues and to affect the dynamical behaviors accordingly.

The Fig.~\ref{Fig6} shows that the resonant frequencies are positively correlated with the imaginary part of the complex eigenvalues, $Im \lambda$, of the Jacobian matrix of the linearized system. As seen in top panel of Fig.~\ref{Fig6}, the $Im \lambda$ plays a role of the lower bound for the resonance frequency $\omega$ and the noise increases the values of resonant frequencies. As briefly discussed in Fig.~\ref{Fig5}, the imaginary part of the complex eigenvalues is related to the rotational angular speed of a deterministic flow in the vicinity of a stable fixed point. According to the subfigures in the upper panel of Fig.~\ref{Fig6}, the resonant frequency of stochastic oscillation is just the angular speed of a deterministic stable spiral when the value of $Im \lambda$ is very large. However, as the value of $Im \lambda$ gets smaller, the resonant frequency gets larger than the angular speed of the deterministic stable spiral. The noise seems to push the system along the deterministic rotational flow rather than to go against it. This tendency is consistently observed in three networks, networks 31, 35, and 59 and under two repression models as presented in the Fig.~\ref{Fig11} in the supplimentary information. Also, as shown in the lower panel of Fig.~\ref{Fig6}, the stochastic oscillation is primarily driven by two causes: the complex eigenvalues of the Jacobian matrix being very close to the imaginary axes and having a larger imaginary part. Those two causes are quantified by two numeric values of proximity and spiralness. The maximum SNR values which are represented by color in heat maps are highly correlated with those two values of proximity as well as spiralness. Proximity is defined as the ratio of the magnitude of noise ($\tilde{B}$ along an eigen-direction to the real negative part of the complex eigenvalues ($Re \lambda$) whereas spiralness is defined as the ratio of the angular speed of a deterministic stable spiral ($Im \lambda$) to the negative real part ($Re \lambda$) which can be thought as an attractive force to the fixed point . Thus the proximity measures the likelihood of making the fixed point stable and pushed the system away from its stable fixed point. The spiralness measures the coherence and amplification of the stochastic oscillations because $Im \lambda$ is very closely related to the resonant frequency and $Re \lambda$ is inversely proportional to the amplitude of the stochastic oscillations, i.e., the peak amplitude of a power spectrum. In the Fig.~\ref{Fig12} in the supplimentary information, one representative network is selected from each performance group: network 59 for blue, network 35 for green, and network 31 for red performance group. The Fig.~\ref{Fig12} clearly shows that for both repression models, the networks belonging to the better performance group have the larger values of proximity and spiralness overall and the larger values of maximum SNR.

\section{Discussion and Conclusions}

This paper is the first exensive comparative study pertaining to a stochastic dynamical behavior from stochastic biochemical reaction networks. Most importantly, we numerically demonstrate the strong correlation between the stochastic behavior and the network architectural properties, namely the coupling patterns of positive and negative feedback loops. We investigate noise-induced oscillation in the networks with only three biochemical species whose reactions are governed by mass action kinetics and with the coupling of positive and negative feedback loops. Modeling a set of so many stochastic biochemical reaction networks by using linear noise approximation and reading the signal-to-noise ratio values from the analytically derived power spectra, we show that all the networks with coupled positive and negative feedbacks are capable to admit the noise-induced oscillations in biologically feasible parameter ranges. Also, using a K-metroid clustering algorithm, we group the sixty-four networks into three performance groups and identify the common network architecture among the networks belonging to the same performance group. We learn that the coupling of negative and positive feedback loops (PNFBL) generally enhance the noise-induced oscillation performance better than the negative feedback loops (NFBL) alone.  However, the performance of PNFBL networks depends on the size of the positive feedback loops (PFBL) relative to that of the NFBL in the networks; the performance of the networks with the bigger PFBL than NFBL is worse than that of the networks with only NFBL.

As shown in the table V in the supplimentary information, we realize that a few networks can generate an unstable fixed point in a very small fraction of the parameter space and this dynamical instability affects the performance of those networks tremendously. The noise-induced oscillation can arise with a significantly high signal-to-noise ratio when the dynamical instability is nearby, whose effect is known as the noisy precursors of the nonlinear  instabilities~\ref{Wiesenfeld}. For the case of proteolysis, all the networks except three networks, networks 10, 31, and 55, have one locally stable fixed point across all parameter ranges. The networks 10, 31, and 55 generate an unstable fixed point with up to 9 pecentage of random sampling from each of four parameter ranges. The networks 10 and 31 outperform the other networks across all parameter ranges whereas the network 55 is in the blue performance group despite having the instabilities inside the parameter sampling ranges. However, the networks from the network architectural group II (networks 14-39) except network 31 don't exhibit any dynamical instability, but are still capable of generating the noise-induced oscillation with the extraordinarily large values of maximum SNR, i.e., up to $maxSNR \sim 10^10$ in the biologically feasible ranges. For the case of transcriptional control, we find that the presence of instabilities of the networks within the chosen parameter range is nicely correlated with the performance of the networks almost without exception. The sample points leading to instabilities goes up to 9.5\% for a few networks, but stay much less than 1\%.

\section{Models and Methods}

\subsection{Biochemical reaction models}
\label{biochem_models}

We consider biochemical reactions in which the chemical species get synthesised and degraded.  The chemical species get synthesised constitutively and their synthesis can be enhanced by another chemical species, called activation. Most often, the activation occurs in a gene regulation where a protein functioning as a transcription factor binds to the promoter of a target gene and enhances the activity of the gene, increasing the production rate of the target protein. The chemical species get degraded spontaneously and they can be negatively regulated by another chemical species, called repression. In this paper, we consider two repression models: repression by proteolysis and repression by transcription control.  Repression by proteolysis can be found in protein-protein leading to proteolysis such as ubiquitin-mediated protein-degradation. Repression by transcription control can be often found in gene reguation where a protein functioning as a repressor binds to a promoter of a target gene and shut off the activity of that gene, decreasing the synthesis of the target protein. 

\begin{figure}[h!]
\begin{center}
\includegraphics[angle=0, height=3.4cm,width=4.0cm]{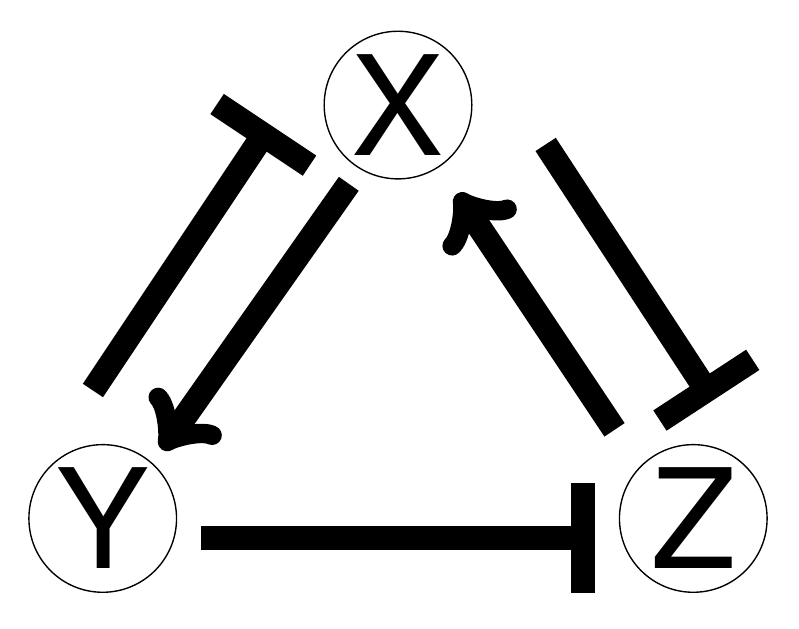}
\end{center}
\caption{NetworkAIAI0I}
\label{example_net}
\end{figure}

\begin{figure}[h!]
\begin{center}
\includegraphics[angle=0, height=8cm,width=15cm]{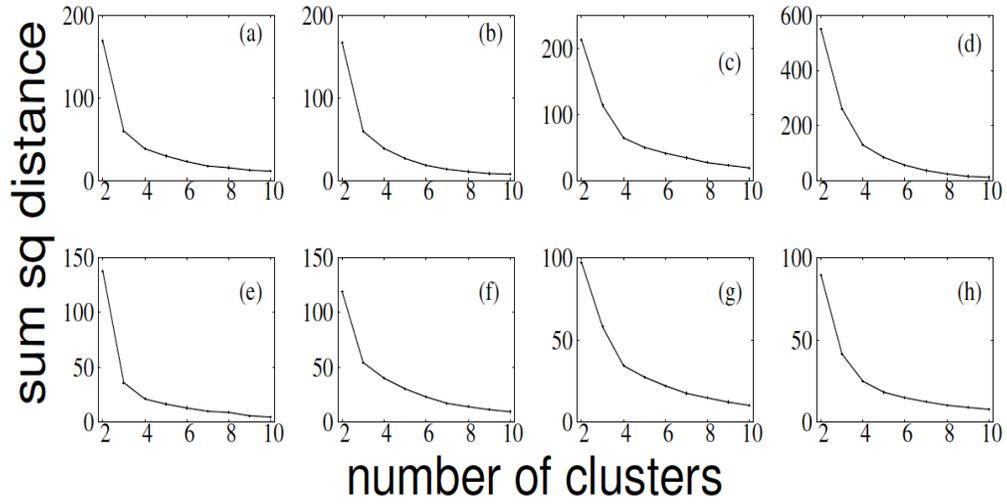}
\caption{Identification of the optimal cluster size by using K-medoids clustering algorithm.}
\end{center}
\label{Fig8}
\end{figure}

\begin{figure}[h!]
\begin{center}
\includegraphics[angle=0, height=22cm,width=18cm]{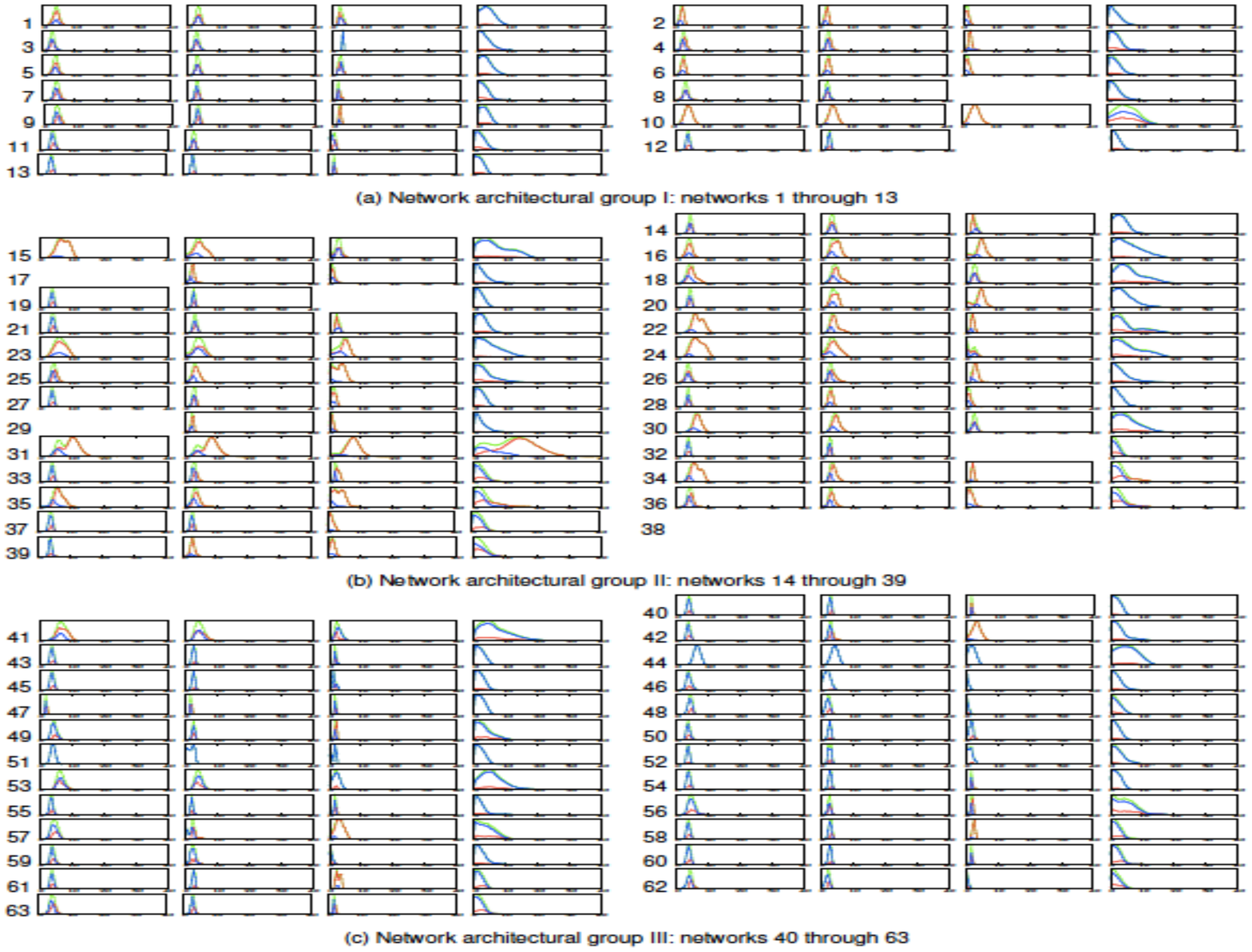}
\end{center}
\caption{Distribution of the Logarithm of the maximum signal-to-noise ratio values for the sixty-three networks with repression by proteolysis. The columns are for different parameter sampling ranges, (a)-(d) from the left to the right. On x-axis is the Log of maximum SNR values whereas on y-axis is the frequency of the maximum SNR values. The blue cuvres denote the distribution of the maximum SNR contributed from the purely real eigenvalues of the jacobian matrix. The red curves depicks the distribution from the complex eigenvalues. Finally, the green curves are the total sum of both distributions from both real and complex eigenvalues.}
\label{Fig9}
\end{figure}

\begin{figure}[h!]
\begin{center}
\includegraphics[angle=0, height=22cm,width=18cm]{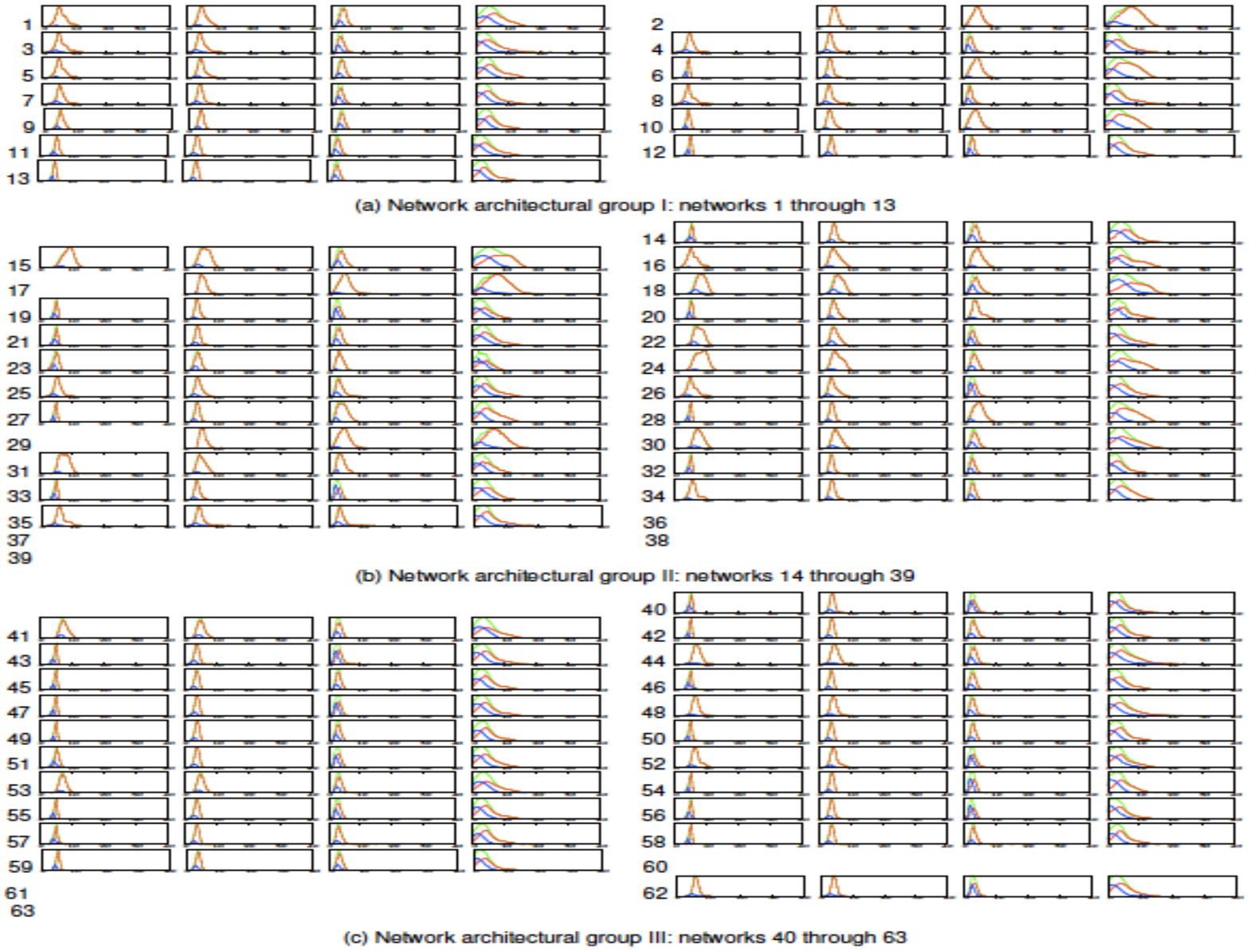}
\end{center}
\caption{Distribution of the Logarithm of the maximum signal-to-noise ratio values for the sixty-three networks with repression by transcriptional control.  The columns are for different parameter sampling ranges, (e)-(h) from the left to the right. On x-axis is the Log of maximum SNR values whereas on y-axis is the frequency of the maximum SNR values. The blue cuvres denote the distribution of the maximum SNR contributed from the purely real eigenvalues of the jacobian matrix. The red curves depicks the distribution from the complex eigenvalues. Finally, the green curves are the total sum of both distributions from both real and complex eigenvalues.}
\label{Fig10}
\end{figure}

\begin{figure}[h!]
\begin{center}
\includegraphics[angle=0, height=15cm,width=15cm]{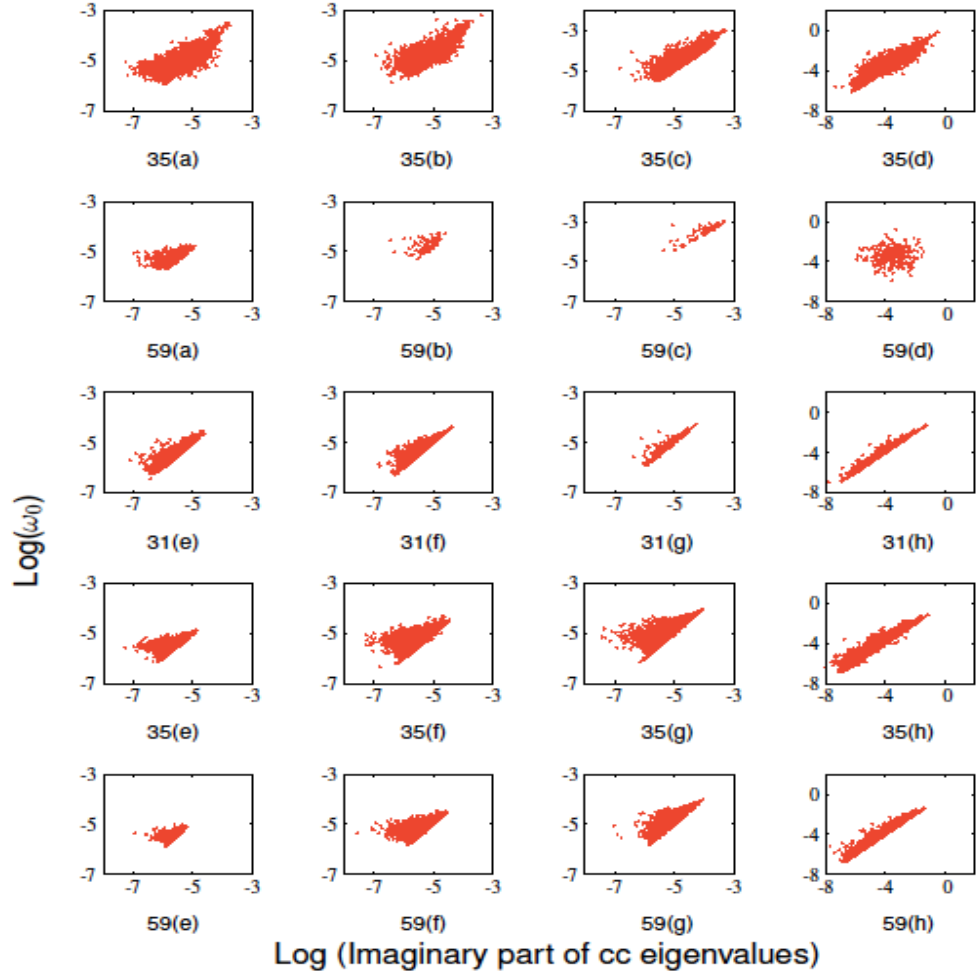}
\end{center}
\caption{Mechanisms of noise-induced oscillation that arises from the complex eigenvalues of the Jacobian matrix. The correlation between the imaginary part of the complex eigenvalues of the Jacobian matrix and the resonant frequency of stochastic oscillation for networks 31, 35, and 59 with both repression models. Each subfigure contains about $10^6$ dots which represent the sets of randomly sampled parameter values.}
\label{Fig11}
\end{figure}

\begin{figure}[h!]
\begin{center}
\includegraphics[angle=0, height=15cm,width=20cm]{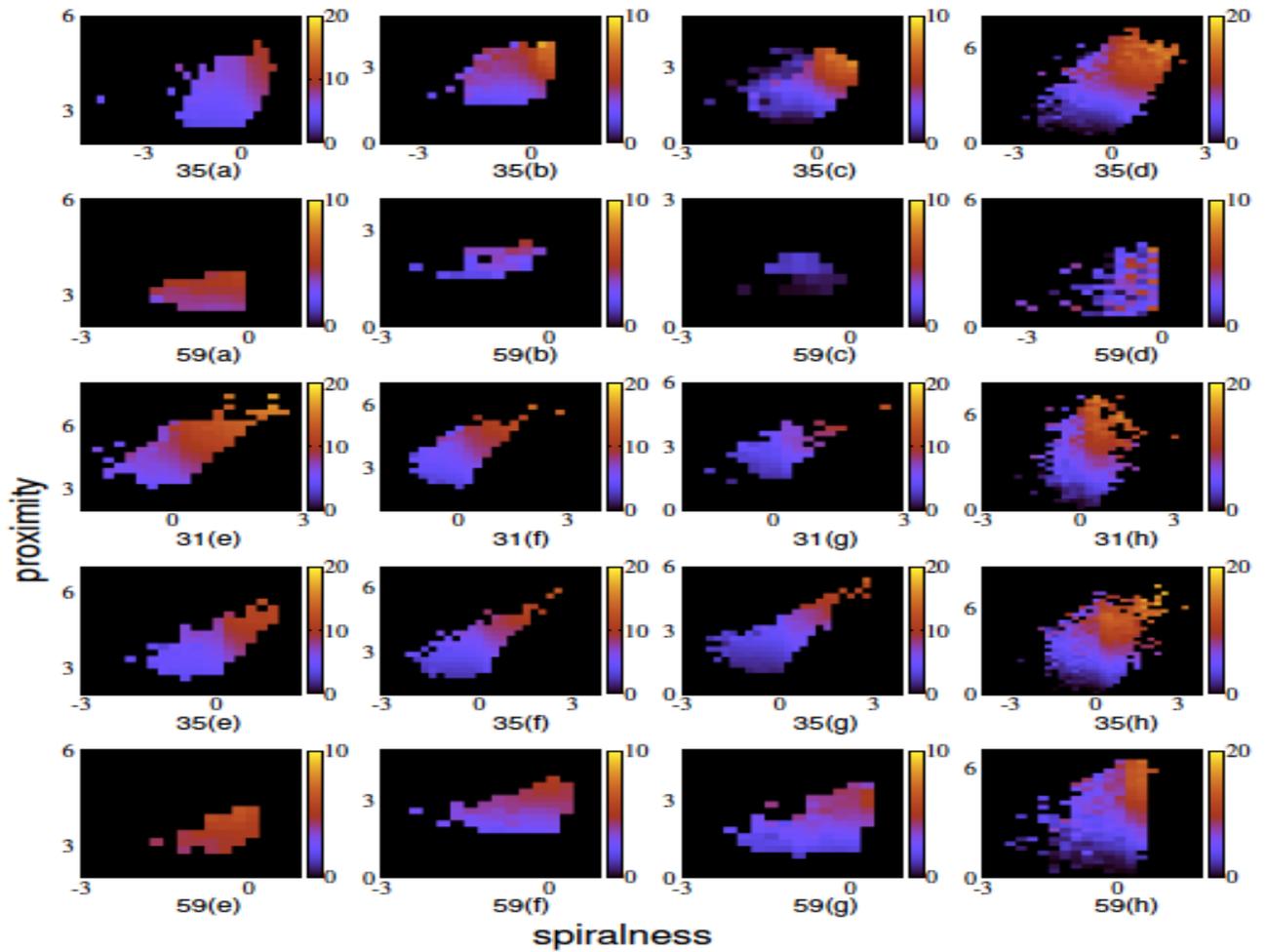}
\end{center}
\caption{Mechanisms of noise-induced oscillation that arises from the complex eigenvalues of the Jacobian matrix. Spiralness versus proximity for the networks 31, 35, snd 59 with both repression models. Both axes are is in logarithmic scale. The colored heat map indicates the logarithmic value of maximum SNR.}
\label{Fig12}
\end{figure}

In this paper, we consider an ensemble of three-node directed graphs with the following constraints: (a) There can exist at most one directed edge from one node to another. (b) When a directed edge is present, it can be either inhibition or activation. In graphical representation, $X \to Y$ is for biochemical species $X$ activating species $Y$, and $X \dashv Y$ is for species $X$ inhibiting species $Y$. (c) We only consider networks in which all three nodes are part of some loop.  Thus networks that have nodes with only in-coming edges or out-going edges or are isolated are not considered.  (d) Self-directed edges are not allowed.  The exhaustive list of three-node directed graphs under our consideration is presented in the Tables~\ref{models} and \ref{models2}.  Table~\ref{models} is for networks with only negative feedback loops while Table~\ref{models2} is for networks with coupled positive and negative feedback loops.  Only topologically distintive networks are allowed.  If two networks are identical after rotation and/or mirroring, then the two networks are topologically identical.  All the graphs can have at most 6 directed edges and the topology of each graph is determined by a distinctive arrangement of the 6 directed edges. Thus, we represent each graph with a string of 6 characters. Each of the 6 characters can be either ``$A$'' for activation, ``$I$'' inhibition, or ``$0$'' when an edge is absent. The first three characters indicate three directed edges going counterclockwise, one from species $X$ to species $Y$, another from species $Y$ to species $Z$, and the last one from species $Z$ to species $X$. The last three characters indicate three directed edges going clockwise, one from species $X$ to species $Z$, another from species $Z$ to species $Y$, and then the last one from species $Y$ to species $X$. For example see Fig. \ref{example_net} for network number 7.  This network consist of three counter-clockwise edges staring from $X$ going to $Y$ (activation), $Y$ to $Z$ (inhibition), and $Z$ to $X$ (activation). There are two clockwise edges in this network one going from $X$ to $Z$ (inhibition) and $Y$ to $X$ (inhibition), and there is no edge going from $Z$ to $Y$. Therefore, we can represent network 7 with a string of `$AIAI0I$'. We represent the rest of the networks in our ensemble by applying the same rules. 

The networks in our ensemble can be categorized into different groups depending on their underlying topological characteristics such as: 1) cyclic negative feedback loops (NFBLs); 2) linear negative feedback loops; 3) cyclic positive-negative interlinked feedback loops (PNFBLs); and, 4) linear positive-negative interlinked feedback loops. The linear feedback loops are made up of only two component feedback loops, which forms a linear chain of three nodes lying in a row. The cyclic feedback loops always involve three component feedbacks as a backbone structure, which form a triangular closed loop. In Table~\ref{models}, we classified networks 1-10 into type I and type II.  Type I NFBLs have ``AIA000'' as the backbone structure while Type II NFBLs have ``III000'' as the backbone structure.  All Type I and Type II NFBLs can be created by adding other links to the backbone structures ``AIA000'' and ``III000'', respectively.  The cyclic PNFBLs can be constructed by adding positive feedback loops in an exhaustive way to the different types of the cyclic NFBL backbone networks. They are termed as the cyclic PNFBLs Type 1 and Type 2 when they are obtained by adding positive feedback loops to NFBLs Type 1 and Type 2, respectively.

In Tables~\ref{models} and \ref{models2}, we further indicate the number and types of feedback loops present in the networks in the form ``$n$N$p$P'', where $n,p=1,2,3$.  Here, $n$ indicate the number of negative feedback loops (N) while $p$ indicate the number of positive feedback loops (P).

\begin{table}[htdp]
\caption{Topologies and classification of networks with negative feedback loops.}
\begin{center}
\begin{tabular}{|c|c|c|}
\hline
Network Identifier  & Network architectural vector & Network classification \\
\hline
1 & AIA000 & Cyclic 3d NFBL (Type 1, 1N) \\
\hline
2 &  III000 & Cyclic 3d NFBL (Type 2, 1N)\\
\hline
3 & AIAI00 &Cyclic 3d NFBL (Type 1, 2N)\\
\hline
4 & AIA00I & Cyclic 3d NFBL (Type 1, 2N)\\
\hline
5 & AIA0A0 & Cyclic 3d NFBL (Type 1, 2N)\\
\hline
6 & III00A & Cyclic 3d NFBL (Type 2, 2N) \\
\hline
7 &AIAI0I & Cyclic 3d NFBL (Type 1, 3N)\\
\hline
8 & AIAIA0 & Cyclic 3d NFBL (Type 1, 3N)\\
\hline
9 & AIA0AI & Cyclic 3d NFBL (Type 1, 3N)\\
\hline
10 & III0AA & Cyclic 3d NFBL (Type 2, 3N)\\
\hline
\hline
11 & AA00II & Linear chain of 2d NFBL, 2N\\
\hline
12 & A0IA0I & Linear chain of 2d NFBL, 2N \\
\hline
13 & 0AIAI0 & Linear chain of 2d NFBL, 2N \\
\hline
\end{tabular}
\end{center}
\label{models}
\end{table}%

\begin{table}[htdp]
\caption{Topologies and classification of networks with both negative and positive feedback loops.}
\begin{center}
\begin{tabular}{|c|c|c|}
\hline
Network Identifier  & Network architectural vector & Network classification \\
\hline
14 &AIA00A & Cyclic 3d NFBL (Type 1, 1P1N)\\
\hline
15 & AIAA00 & Cyclic 3d NFBL (Type 1, 1P1N)\\
\hline
16 & AIA0I0 & Cyclic 3d NFBL (Type 1, 1P1N)\\
\hline
17 & IIII00 & Cyclic 3d NFBL (Type 2, 1P1N)\\
\hline
\hline
18 &AIAA0A & Cyclic 3d NFBL (Type 1, 2P1N)\\
\hline
19 &AIAI0A & Cyclic 3d NFBL (Type 1, 1P2N) \\
\hline
20 &AIA0IA &  Cyclic 3d NFBL (Type 1, 2P1N)\\
\hline
21 & AIA0AA & Cyclic 3d NFBL (Type 1P2N)\\
\hline
22 & AIAA0I & Cyclic 3d NFBL (Type 1, 1P2N)\\
\hline
23 & AIAAI0 & Cyclic 3d NFBL (Type 1, 2P1N)\\
\hline
24 & AIAAA0 & Cyclic 3d NFBL (Type 1, 1P2N)\\
\hline
25 &  AIA0II & Cyclic 3d NFBL (Type 1, 1P2N)\\
\hline
26 & AIAII0 & Cyclic 3d NFBL (Type 1, 1P2N)\\
\hline
27 & IIII0A & Cyclic 3d NFBL (Type 2, 1P2N) \\
\hline
28 & IIIIA0 & Cyclic 3d NFBL (Type 2, 1P2N) \\
\hline
29 &  IIII0I & Cyclic 3d NFBL (Type 2, 2P1N)\\
\hline
\hline
30 & AIAAAA & Cyclic 3d NFBL (Type 1, 3P2N)\\
\hline
31 & AIAAIA & Cyclic 3d NFBL (Type 1, 3P2N)\\
\hline
32 & AIAIAA & Cyclic 3d NFBL (Type 1, 1P4N)\\
\hline
33 & AIAIIA & Cyclic 3d NFBL (Type 1, 3P2N)\\
\hline
34 & AIAAAI & Cyclic 3d NFBL (Type 1, 1P4N)\\
\hline
35 & AIAAII & Cyclic 3d NFBL (Type 1, 3P2N)\\
\hline
36 & AIAIII & Cyclic 3d NFBL(Type 1, 1P4N )\\
\hline
37 & IIIIIA & Cyclic 3d NFBL (Type 2, 3P2N)\\
\hline
38 & IIIIII& Cyclic 3d NFBL (Type 2, 3P2N)\\
\hline
39 &  IIIAAA & Cyclic 3d NFBL (Type 2, 1P4N)\\
\hline
\hline
40 & AA00AI & Linear chain of 2d NFBL/PFBL (1P1N) \\
\hline
41 & IA00AA & Linear chain of 2d NFBL/PFBL (1P1N)\\
\hline
42 & AI00II & Linear chain of 2d NFBL/PFBL (1P1N)\\
\hline
43 &  II00IA & Linear chain of 2d NFBL/PFBL (1P1N)\\
\hline
\hline
44 & AAA00I &  Cyclic 3d PFBL + 2dNFBL (1P1N) \\
\hline
45 & AII00I & Cyclic 3d PFBL + 2dNFBL (1P1N) \\
\hline
46 & AIIA00 & Cyclic 3d PFBL + 2dNFBL (1P1N) \\
\hline
47 & AII0A0 & Cyclic 3d PFBL + 2dNFBL (1P1N) \\
\hline
\hline
48 & AAAI0I & Cyclic 3d PFBL + linear chain of 2dNFBL (1P2N) \\ 
\hline
49 & AII0AI & Cyclic 3d PFBL + linear chain of 2dNFBL (1P2N) \\
\hline
50 & AIIA0I & Cyclic 3d PFBL + linear chain of 2dNFBL (1P2N) \\
\hline
51 & AIIAA0 & Cyclic 3d PFBL + linear chain of 2dNFBL (1P2N) \\
\hline 
52 & AAAI0A & Cyclic 3d PFBL + linear chain of 2dNFBL (2P1N) \\
\hline 
53 & AAAIA0 &  Cyclic 3d PFBL + linear chain of 2dNFBL (2P1N)\\
\hline 
54 & AIIIA0 & Cyclic 3d PFBL + linear chain of 2dNFBL (2P1N)\\
\hline 
55 & AIII0I & Cyclic 3d PFBL + linear chain of 2dNFBL (2P1N) \\
\hline 
56 & AII0AA & Cyclic 3d PFBL + linear chain of 2dNFBL (2P1N) \\
\hline
57 & AIIAI0 & Cyclic 3d PFBL + linear chain of 2dNFBL (2P1N)\\
\hline 
58 & AII0II & Cyclic 3d PFBL + linear chain of 2dNFBL (2P1N)\\
\hline 
59 & AIIA0A & Cyclic 3d PFBL + linear chain of 2dNFBL (2P1N)\\
\hline
\hline
60 & AIIIAI &  Cyclic 3d PFBL + 2dNFBL (3P2N)\\
\hline
61 & AIIAII &  Cyclic 3d PFBL + 2dNFBL (3P2N)\\
\hline
62 & AIIAAA &  Cyclic 3d PFBL + 2dNFBL (3P2N)\\
\hline
63 & AIIAAI &  Cyclic 3d PFBL + 3d NFBL + 2dNFBL (3P2N)\\
\hline
\end{tabular}
\end{center}
\label{models2}
\end{table}%

\subsection{Master Equations}
In order to study the stochastic dynamics of a network, we start by writing down the chemical master equation that describes the time evolution of the states of the network of chemical reactants. We denote the state of the network as a vector $\tilde{S} = (X,Y,Z) $ that represent the current populations of the three chemical species of the network.  We also denote the joint probability distribution of  the state of the network at a particular time $t$ as $P(\tilde{S};t)$ and the transition rate from state $\tilde{S'}$ to state $\tilde{S}$ as $T(\tilde{S} | \tilde{S'})$.  This allows us to formally write the master equation governing the time evolution of $P(\tilde{S};t)$ as 
\begin{equation}
\frac{ \partial P(\tilde{S};t)}{ \partial t} = \sum_{\tilde{S'} } T(\tilde{S} | \tilde{S'}) P(\tilde{S'};t) -  \sum_{\tilde{S} } T(\tilde{S'} | \tilde{S}) P(\tilde{S};t)
\end{equation}
To mathematically model the transition rates among the states, we make the following assumptions: (a) any given node is constitutively synthesized, which depends on the system size $\Omega$, (b) any given node is degraded at a rate which depends on its current abundance, (c) activation is proportional to the abundance of the activating species, (d) for inhibitory interaction, we consider two models as discussed in Sec. \ref{biochem_models}, repression by proteolysis and repression by transcription control; and (e) the corresponding stochastic chemical reaction is a Markovian process. Based on these assumptions, we can specify all the allowed transitions among different states.  In Table~\ref{T-rates}, we show the transition rates of synthesis and degradation associated with all the possible chemical reactions that can occur on a single node $X$. In Table~\ref{T-rates}, we have shown the transition rates for both repression by proteolysis and repression by transcritpion control discussed in Sec. \ref{biochem_models}.  These transition rules combined with the particular network topology fully determine the mathematical formulation of the chemical master equations.

\begin{table}[htdp]
\caption{The transition rates for synthesis and degradation associated with each chemical reaction. ${k_i}$'s are the kinetic rate constants.}
\begin{center}
\begin{tabular}{|c|c|c|}
\hline
Reactions  & & Transition rate \\
\hline
$\emptyset \to X$ & constitutive synthesis & $k_1$ \\
\hline
$X \to \emptyset $ & spontaneous degradation &  $k_2 X $   \\
\hline
$Y \to X$ & activation & $k_3 Y$ \\
\hline
$Y \to X$,  $Z \to X$ & activation & $k_3(Y+Z) $ \\
\hline
$Y \dashv X$ & proteolysis  & $k_4 XY$ \\
\hline
$Y \dashv X$,  $Z \dashv X$ & proteolysis & $k_4 (XY + XZ)$ \\
\hline
$Y \to X,Z \dashv X$ &activation, transcription control & $(k_1+k_3Y)e^{-k_5Z}$ \\
\hline
$Z \dashv X$ & no activation,transcription control & $k_1e^{-k_5Z}$ \\
\hline
$Y \dashv X$,  $Z \dashv X$ & no activation,transcription control & $k_1e^{-k_5Y-k_6Z}$ \\
\hline
\end{tabular}
\end{center}
\label{T-rates}
\end{table}%

The master equation can be rewritten by using the step operator notation in the following way:
\begin{equation}
\frac{ dP(\tilde{S})}{dt} = \sum_{S_i=X,Y,Z} [E_{S_i}^{-}-1] T(\tilde{S'}|\tilde{S}) P(\tilde{S})   + \sum_{S_i=X,Y,Z}  [E_{S_i}^{+}-1] T(\tilde{S}|\tilde{S'}) P(\tilde{S'}) 
\label{M-eq}
\end{equation}
where the step operators for species X, $E_X^{+}$ and $E_X^{-}$, are defined by
\begin{equation}
E_X^{\pm} f (X, Y, Z)  = f (X \pm 1, Y, Z)  = \sum_{m = 0}^{\infty} (-1)^{\pm m} \frac{\partial^m}{\partial {X}^m}  f (X, Y, Z)
\end{equation}
where $f(X,Y,Z)$ is an analytic function.  Operators $E_Y^{\pm}$ and $E_Z^{\pm}$ are similarly defined. Thus, the operation of the step operator is to either raise or lower the number of a node by 1.

\subsection{System size expansion}
The form of the chemical master equation does not allow the analytic solution in general due to the nonlinearity. One possible way to deal with this problem is to use the system size expansion method introduced by van Kampen. It allows us to expand the master equation systematically  in terms of the system size $\Omega$. Thereby, we can obtain the approximate solutions to the leading orders of the large system size. 

Here, we sketch the standard procedure of the system size expansion briefly following Ref.~\cite{Kampen}. Interested readers should refer to Ref.~\cite{Kampen} for more detail. The main idea of the system size expansion assumes that the abundance of chemical species can be decomposed into its macroscopic mean value and the fluctuation around the mean, both of which depends on the system size $\Omega$. For example, we can write the number of chemical species $S_i$ as
\begin{equation}
S_i = \Omega \phi_i + \Omega^{1/2} {\xi}_i 
\label{Xtox}
\end{equation}
where the index $i$ runs over all the chemical species, i.e., $S_i = {X,Y,Z}$. The first term $\Omega \phi_i$ describes the macroscopic mean value of $S_i$. The second term  $\Omega^{1/2} \xi_{i}$ is the gaussian fluctuations of $S_i$ around its mean. Then, we can rewrite the joint probability distribution $P(\{S_i\}; t)$ in terms of the set of stochastic variables of $\xi_i$ instead of $S_i$. Upon such change of variables, the time derivative of the joint probability distribution transforms as
\begin{equation}
\frac{\partial P(\tilde{S})}{\partial t} =  \frac{\partial \Pi(\tilde{\xi})}{\partial t} - \sum_{\xi_i} \Omega^{1/2} \frac{d\phi_i}{dt} \frac{\partial \Pi(\tilde{\xi})}{\partial \xi_i} 
\end{equation}
Accordingly, the step operators are also rewritten in terms of the new variable.
\begin{equation}
E_{S_i}^{\pm} \to E_{\xi_i}^{\pm} =\sum_{m=0}^{\infty} (-1)^{\pm m} \Omega^{-\frac{m}{2}} \frac{\partial^m}{\partial \xi_i^m}.  
\label{Step-OP}
\end{equation}
Plugging Eq.~(\ref{Xtox}) through Eq.~(\ref{Step-OP}) in Eq.~(\ref{M-eq}), we can expand the master equation systematically in the order of the system size $\Omega$. To the leading order of $\Omega$ expansion $(\Omega^{1/2})$, we obtain a set of the macroscopic deterministic equations. In the next leading order of $\Omega$ expansion $(\Omega^0)$, we obtain the linear Fokker-Plank equation.
\begin{equation}
\frac{\partial \Pi}{\partial t} = - \sum_{\xi_i} \frac{ \partial}{\partial \xi_i  }C(\xi_i) \Pi + \sum_{\xi_i,\xi_j} \frac{1}{2} B_{ij} \frac{\partial^2 \Pi}{\partial \xi_i \partial \xi_j }
\label{FP-eq}
\end{equation}
where $C(\xi_i) \equiv \sum_{j} J_{ij} \xi_j$. $J_{ij}$ denotes the $(i,j)^{th}$ element of the Jacobian matrix $J$ of the macroscopic rate equations and  $B_{ij}$ denotes the $(i,j)^{th}$ element of the noise covariance matrix $B$.

\subsection{Power spectrum}
\label{power_spectrum}
We can recast the linear Fokker-Plank equation into the mathematically equivalent form of Langevin equations.
\begin{equation}
\frac{d \xi_i(t)}{dt} = \sum_{j} J_{ij} \xi_j(t) + \eta_i(t).
\label{langevin}
\end{equation}
where $\langle\eta_i(t) \eta_j(t')\rangle = B_{ij} \delta(t-t')$. The matrix elements $J_{ij}$ and $B_{ij}$ are evaluated at the fixed points of the macroscopic deterministic equations. 

In order to examine the existence of the stochastic ocillations, we can study the power spectrum of the stochastic variable $\xi_i(t)$. Namely, 
\begin{equation}
 P_i(\omega) = \langle |\tilde{\xi_i}(\omega)|^2\rangle = \sum_{j,k} \Phi_{ij}^{-1}(\omega) B_{jk} (\Phi_{ki}^{-1})^{\dagger}(\omega) 
\end{equation}
where $\tilde{\xi_i}(\omega)$ denotes the Fourier transform of $\xi_i(t)$ and $\Phi_{ij} \equiv -i\omega \delta_{ij} - J_{ij}$.

For our case, $J$ and $B$ are 3 by 3 matrices.  The explicit expression of the power spectrum for species $i$, in terms of matrix elements of $J$ and $B$, is given by 
\begin{equation}
P_i(\omega) =\frac{ \beta_{1i}\omega^4 + \beta_{2i} \omega^2 + \beta_{3i} }{\omega^6 + \alpha_1\omega^4 + \alpha_2\omega^2 + \alpha_3}
\label{Pw}
\end{equation}
where the coefficients  in the numerator are: $\beta_{1i}= B_{ii}$, $\beta_{2i} = B_{ii}[ (Tr[J]-J_{ii})^2 -2M_{ii} ] + B_{i+1,i+1}J_{i,i+1}^2 +B_{i+2,i+2}J_{i,i+2}^2$, and $\beta_{3i}= B_{ii}M_{ii}^2 + B_{i+1,i+1}M_{i+1,i}^2+ B_{i+2,i+2}M_{i+2,i}^2$ and $M_{ij}$ denotes the minor matrix of $J_{ij}$. The index of the matrices obeys the following conventions: $J_{i,j}=J_{i+3,j+3}$, $B_{i,j}=B_{i+3,j+3}$, and $M_{i,j} = M_{i+3,j+3}$ (cyclic...).  The coefficients in the denominator are: $\alpha_1 = Tr[J]^2-2(J_{11}J_{22}+J_{22}J_{33}+J_{33}J_{11} - J_{12}J_{21} -J_{23}J_{32} - J_{31}J_{13}) $, $\alpha_2 = (J_{11}J_{22}+J_{22}J_{33}+J_{33}J_{11} - J_{12}J_{21} -J_{23}J_{32} - J_{31}J_{13})^2 - 2Tr[J]Det[J]$, and $\alpha_3 = (Det[J])^2$.

\subsection{Sampling methods and root calculation}
\label{sampling-and-root}
The networks we consider in this work are stable in the positive real domain in the parameter space, i.e., $\{k_{i}\} \in R^{+}$ where $\{k_i\}$ are the kinetic rate constants shown in Table~\ref{T-rates}. In order to demonstrate the existence of the stochastic amplified oscillations, we sample $10^6$ sets of kinetic rate constants uniformly and randomly from the logarithmically scaled intervals by using the unconstrained Monte Carlo sampling technique.  For both models that incorporate repression by proteolysis and repression by tracription control, we sample kinetic rate constants from biologically feasible range and uniform range.  For the biologically feasible range, we sample constitutive synthesis rates from the interval $[10^{-7}, 10^{-5})$, and enhanced synthesis and degradation rates from the interval $[10^{-6}, 10^{-4})$.  Further, repression rates were sampled from three separate intervals.  For repression by proteolysis, $[10^{-7}, 10^{-5})$, $[10^{-5}, 10^{-3})$ and $[10^{-3}, 10^{-1})$.  For repression by transcription control, $[10^{-3}, 10^{-1})$, $[10^{-1}, 10^1)$ and $[10^1, 10^3)$.  For the uniform range, we sample constitutive synthesis rates, enhanced synthesis rates and degradation rates from the interval $[10^{-7}, 10^{-1})$.  For repression by proteolysis $[10^{-7}, 10^{-1})$.  For repression by transcription control $[10^{-3}, 10^3)$.

\begin{table}[htdp]
\caption{Sampling intervals for the kinetic rate constants.}
\begin{center}
\begin{tabular}{|c|c|c|c|c|}
\hline
Reaction  & bio interval  & bio interval & bio interval & uniform interval \\
\hline
Constitutive synthesis & $[10^{-7}, 10^{-5})$ & $[10^{-7}, 10^{-5})$ & $[10^{-7}, 10^{-5})$ & $[10^{-7}, 10^{-1})$ \\
\hline
Enhanced synthesis & $[10^{-6}, 10^{-4})$ & $[10^{-6}, 10^{-4})$ & $[10^{-6}, 10^{-4})$ & $[10^{-7}, 10^{-1})$ \\
\hline
Degradation & $[10^{-6}, 10^{-4})$ & $[10^{-6}, 10^{-4})$ & $[10^{-6}, 10^{-4})$ & $[10^{-7}, 10^{-1})$ \\
\hline
Repression by proteolysys & $[10^{-7}, 10^{-5})$ & $[10^{-5}, 10^{-3})$ & $[10^{-3}, 10^{-1})$ & $[10^{-7}, 10^{-1})$ \\
\hline
Repression by transcription control & $[10^{-3}, 10^{-1})$ & $[10^{-1}, 10^{1})$ & $[10^{1}, 10^{3})$ & $[10^{-3}, 10^{3})$ \\
\hline
\end{tabular}
\end{center}
\label{k_intervals}
\end{table}

We need to know the roots of the deterministic system to calculate the power spectra using Eq. (\ref{Pw}).  For some networks the roots of the deterministic system can be obtained analytically.  For networks for which the steady state of the deterministic system can not be obtained analytically, we used the bisection method.  We left out few networks for which it becomes impossilbe to get steady state values either by analytical expressions or by using bisection method, e.g., when two or more equations are coupled. Since the system size expansion can only be used for the system that has a single stable steady state, for proteolysis, we check the coefficients of the polynomial equations to makes sure that we have a single positive root for a given network for a given set of randomly sampled kinetic rate constants. We resample when we encounter either a negative or a very large root.  This is because negative and very large steady states are biologically unfeasible.  Thus by resampling we exclude such parts of domian. Maximum limit was set to a very high value ($10^{15}$). After obtaining a single positive state, we check for the stability of the dynamical system numerically as described in Sec. \ref{lin-stability}.  If the system is unstable, we resample $k_i$'s from the parameter space. When the root calculation fails in the bisection method, we set $SNR$ values to zero and do not resample.

\subsection{SNR calculation}
The SNR (signal to noise ratio) is commonly defined by the power spectrum peak height over its relative width.  For species $i$ it is thus
\begin{equation}
SNR_i = \frac{P_i(\omega_o)}{\Delta \omega/ \omega_o}
\end{equation}
where $P_i(\omega_o)$ and $\omega_o$ denote the peak height and the peak frequency, respectively, for species $i$. The $\Delta \omega$ defines the so-called full width at half maximum (FWHM).  
Our calculation of the power spectrum is based on the analytical expressions discussed in section \ref{power_spectrum}. It is, however, not straightforward to obtain the analytic expression for SNR because SNR calculation requires the solution of fourth order polynomial equation given by $dP_i(z)/dz = 0$ where $z = \omega^2$ and the determination of $\Delta \omega$ around $\omega_o$ is also non trivial. Therefore, we use a numerical algorithm to calculate the SNR in the following way. 
First, we numerically identify the occurrence of the maxima, if it exist, of $P_i(\omega)$ in Eq. (\ref{Pw}) for a given set of kinetic rate constants and fixed points of the system.  We locate the peak frequency at $\omega = \omega_o$ for which $P_i(\omega_o)$ is the maximum value.  Next, if the peak frequency $\omega_o$ exists, we search for the two neighboring frequencies, i.e., $\omega_1 (< \omega_o )$ and $\omega_2 (>\omega_o)$ that satisfy $P_i(\omega_1) = P_i(\omega_2) = P_i(\omega_o)/2$. Then, we can determine $\Delta \omega$ from the difference of two frequencies, i.e., $\Delta \omega = |\omega_2 - \omega_1|$.        
 
\subsection{Robustness vs Prominence}
In this work, the robustness of a networks is defined as the size of the domain of the parameter space that generates stochastic amplified oscillations, i.e., SNR$ \ge 1$, whereas, the prominence is defined as the average value of SNR within such domain for that network.  It is to be noted that we only consider $\{k_i\}$ domain in which we do not have multiple roots, negative roots, very large roots and unstable roots.  As discussed in Sec. \ref{sampling-and-root}, when the bisection method fails, we set SNR values to zero and do not resample.  This gives us lower bounds on robustness.  We measure the size of the domain of the parameter space that generates stochastic amplified oscillations by counting the data sets that give SNR $ \ge 1$ out of total sampling points we accept.  For species $i$, we calculate the average value of SNR within the domain by integrating the probability distribution of SNR, i.e.,
\begin{equation}
\langle {SNR_i} \rangle = \frac { \int SNR_i ~P_i(SNR_i) ~\Theta(SNR_i - 1) ~d(SNR_i) } { \int P_i(SNR_i) ~\Theta(SNR_i - 1) ~d(SNR_i) }.
\end{equation}
The step function $\Theta(x)$ is inserted to restrict the integral range only for SNR$\ge 1$.  In numerical simulations, $(\sum\limits_{j,SNR_{i,j}\ge 1} SNR_{i,j})/D_i$.  Here $j$ is the data point and $D_i$ is the counts of the data sets with $SNR_i\ge 1$ for species $i$ in the network.

\subsection{Eigenvalue calculation}

The dimensions of Jacobian matrices for all the networks considered in this parper are 3-by-3 and the corresponding third order characteristic polynomials can have only real coefficients.  The three eigenvalues of the Jacobian matrix can be either all real or a combination of a real eigenvalue and a complex conjugate pair.  Since all the networks we consider exhibit stable deterministic dynamics in the parameter range of our interest, the real parts of all three eigenvalues are necessarily negative.  We used cubic polynomial formulae for obtaining the eigenvalues of 3-by-3 Jacobian matrices \cite{Murray}. We reject any samples for which numerical accuracy is a problem, e.g., for some Jacobian matrices the numerical variables become so small that the calculation using double precision is not adequate and may need higher precision.  
Apart from looking at the imaginary part of eigenvaules, we can use discriminant of the matrix to determine if the eigenvalues are complex or real.  Discriminant $< 0$ means that we have a complex conjugate pair, otherwise, all three eigevalues are real. 

\subsection{Linear stability}
\label{lin-stability}
The proof of the stability for the deterministic system amounts to the examination of  $Re(\lambda)<0$, where $\lambda$ denotes the eigenvalues of the Jacobian matrix, which are the roots of the cubic characteristic polynomials, i.e., $ det(J - I \lambda) = 0$:
\begin{equation}
\lambda^{3}-Tr(J)\lambda^2+(A_{11}+A_{22}+A_{33})\lambda - det(J) = 0.
\end{equation}
Here $Tr(J)$ representing the trace of the Jacobian, while $A_{ij}$ is the cofactor of the $i^{th}$ row and $j^{th}$ column. For all the eigenvalues to have $Re(\lambda) <0$, the Routh-Hurwitz conditions for a third order characteristic equation, $Tr(J)<0, ~~ det(J)<0, ~~det(J)-(A_{11}+A_{22}+A_{33})Tr(J)>0$ should be satisfied.  We used Routh-Hurwitz condition to numerically check the stability condition.

\subsection{$K$-medoids clustering}
We used $k$-medoids algorithm to cluster the data points.  $K$-medoids is a partitioning technique of clustering a data set into a given number of clusters.  We chose this method over $K$-means clustering because it is more robust to noise and outliers.  We used the squared Euclidean distance as our distance measuer.  A medoid of a cluster is the most centrally located data point of the cluster.  The most common realisation of $K$-medoid clustering is the Partitioning Around Medoids algorithm which is also the method we used in this paper.  Since the attainment of the minima is not guaranteed, we made several trials with different initial conditions and chose the clustering that gave least squared Euclidean distance.

\subsection{Calculation of Hamming distance for the identification of common network architecture}
We use the Hamming distance measure to find the similarity between any two given networks. To find the Hamming distance between two given networks, in both the networks we assign the value of $+1$ to an edge when it is activation, $-1$ to an edge when it is repression and $0$ when the edge is absent.  We then overlap the two networks and calculate the Hamming distance as the sum of the absolute values of the differences of the values of the edges that point in the same direction.  While calculating the Hamming distance between two given networks, we also consider the mirror images of the networks, and the rotations of the networks and their mirror images and choose the Hamming distance which is least after considering these transformations.

To calculate the common network architecture of each cluster, one-by-one, we consider each network in the cluester as a reference network of the cluster.  We then find the sums of the Hamming distances of the reference networks from the rest of the networks in its cluster.  We choose the minimum of these sums of Hamming distances and the configuration corresponding to this minimal value.  Note that the configuration may contin the rotations and mirror images of the networks because we also consider these transformations while calculating the Hamming distance.  The common architecture is the average architecture over all the networks corresponding to this configuration.  In the common architecture, a positive value of an edge means that the edge is activation while a negative value means the edge is repression.  The magnitude of the average edge values indicate how common the edge is within the cluester.  Magnitudes close to $1$ means the edge is very common inside the cluester while a value close to $0$ means the edge is uncommon.

\newpage 

\section{Supplimentary Information}

\begin{table}[htdp]
\caption{Percentage of the samples yielding locally unstable fixed points}
\begin{tabular}{|l|l|l|l|l|l|l|l|l|}
\hline
Net & A & B & C & D & E & F & G & H \\
\hline
1 & 0.0 & 0.0 & 0.0 & 0.0 &  0.0 & 0.0 & 0.0 & 0.00011998 \\
2 & 0.0 &  0.0 & 0.0 &  0.0 & 0.0 &  0.00000699 & 0.08844374 &  0.01895958  \\
3 &  0.0 & 0.0 & 0.0 & 0.0 & 0.0 & 0.0 & 0.0 & 0.00003599 \\
4  & 0.0 & 0.0 & 0.0 &  0.0 & 0.0 &  0.0 & 0.0 & 0.00001099 \\
5 & 0.0 & 0.0 & 0.0 & 0.0 & 0.0 & 0.0 & 0.0 & 0.00011198  \\
6 & 0.0 & 0.0 & 0.0 &  0.0 & 0.0 & 0.00001099 & 0.09177933 & 0.02205558 \\
7 & 0.0 & 0.0 & 0.0 & 0.0 & 0.0 &  0.0 & 0.0 & 0.00000499 \\
8  &  0.0 & 0.0 & 0.0 & 0.0 & 0.0 & 0.0 & 0.0 & 0.00004199  \\
9 & 0.0 & 0.0 &  0.0 & 0.0 & 0.0 & 0.0 & 0.0 & 0.00001199 \\
10  & 0.00129531 & 0.00127537 & 0.08899183 &  0.01733422 & 0.0 & 0.00002899 &  0.09498658 & 0.02643141 \\
11 &  0.0 & 0.0 & 0.0 & 0.0 & 0.0 & 0.0 & 0.0 & 0.0 \\
12 & 0.0 & 0.0 & 0.0 & 0.0 & 0.0 & 0.0 & 0.0 & 0.0 \\
13 & 0.0 &  0.0 & 0.0 & 0.0 & 0.0 & 0.0 & 0.0 & 0.0  \\
14 & 0.0 & 0.0 & 0.0 & 0.0 & 0.0 & 0.0 & 0.0 & 0.00000699  \\
15 & 0.0 & 0.0 &  0.0 & 0.0 & 0.00159943 & 0.00020395 &  0.0 & 0.00204281 \\
16 & 0.0 & 0.0 & 0.0 & 0.0 & 0.0 &  0.00014497 & 0.00020495 &  0.0008313 \\
17 &  0.0 & 0.0 &  0.0 & 0.0 & 0.0 & 0.00000699 &  0.04777956 & 0.011636 \\
18  & 0.0 & 0.0 & 0.0 &  0.0 &  0.0 & 0.0 &  0.0 & 0.00051673 \\
19 &  0.0 & 0.0 & 0.0 & 0.0 &  0.0 & 0.0 &  0.0 & 0.00000099 \\
20 & 0.0 & 0.0 & 0.0 &  0.0 & 0.0 & 0.00002099 & 0.00004899 &  0.00027092 \\
21 & 0.0 & 0.0 & 0.0 & 0.0 & 0.0 & 0.0 & 0.0 &  0.00000799 \\
22  &  0.0 & 0.0 & 0.0 & 0.0 &  0.00003899 & 0.0 & 0.0 & 0.00020095 \\
23 & 0.0 & 0.0 &  0.0 & 0.0 & 0.0 & 0.00000099 &  0.00000299 &  0.00001799  \\
24 & 0.0 & 0.0 &  0.0 &  0.0 & 0.0005447 & 0.00004699 &  0.0 & 0.00103792 \\
25 & 0.0 &  0.0 & 0.0 & 0.0 & 0.0 & 0.00010998 & 0.00135516 & 0.00127138 \\
26 &  0.0 &  0.0 & 0.0 & 0.0 & 0.0 & 0.00002199 & 0.00004699 & 0.00040583 \\
27 & 0.0 & 0.0 &  0.0 & 0.0 & 0.0 &  0.00002599 & 0.02537057 & 0.00872225 \\
28  & 0.0 &  0.0 & 0.0 &  0.0 & 0.0 & 0.00000899 & 0.04864288 & 0.01375515 \\
29 & 0.0 & 0.0 &  0.0 &  0.0 & 0.0 & 0.00000499 &  0.0099 & 0.00355332  \\
30 & 0.0 & 0.0 &  0.0 & 0.0 & 0.0 & 0.0 & 0.0 & 0.00021195 \\
31 &  0.05820748 & 0.04891976 & 0.04897584 & 0.04896589 & 0.00010198 & 0.00009599 & 0.00002599 &  0.00012198 \\
32  & 0.0 & 0.0 &  0.0 & 0.0 &  0.0 &  0.0 & 0.0 &  0.00000099 \\
33 &  0.0 & 0.0 & 0.0 &  0.0 &  0.00001299 &  0.00062361 &  0.00517408 & 0.00039784 \\
34  & 0.0 & 0.0 & 0.0 &  0.0 & 0.0 & 0.0 & 0.01601823 & 0.09363806 \\
35 & 0.0 & 0.0 & 0.0 & 0.0 & 0.00000399 & 0.00004599 & 0.01718066 & 0.09201023  \\
36 & 0.0 & 0.0 & 0.0 & 0.0 & - &  - &  - & -  \\
37  & 0.0 & 0.0 & 0.0 & 0.0 & - & - & - & - \\
38 & - &  - &  - &  - & - &  - &  - & - \\
39 & 0.0 & 0.0 & 0.0 & 0.0 & - &  - & - &  - \\
\hline
\end{tabular}
\end{table}

\begin{table}
\begin{tabular}{|l|l|l|l|l|l|l|l|l|}
\hline
Net & A & B & C & D & E & F & G & H \\
\hline
40 & 0.0 & 0.0 & 0.0 & 0.0 & 0.0 & 0.0 & 0.0 & 0.0  \\
41 & 0.0 & 0.0 & 0.0 & 0.0 & 0.0 & 0.0 & 0.0 & 0.0 \\
42 & 0.0 & 0.0 & 0.0 &  0.0 & 0.0 & 0.00000499 & 0.00081034 & 0.00059464 \\
43 &  0.0 &  0.0 & 0.0 & 0.0 & 0.0 &  0.0 & 0.00000999 & 0.00011998 \\
44 & 0.0 & 0.0 & 0.0 &  0.0 & 0.0 & 0.0 &  0.0 & 0.0 \\
45 & 0.0 & 0.0 &  0.0 & 0.0 & 0.0 & 0.0 & 0.0 & 0.0  \\
46 & 0.0 & 0.0 &  0.0 & 0.0 & 0.0 & 0.0 & 0.0 &  0.0  \\
47 &  0.0 & 0.0 & 0.0 & 0.0 & 0.0 & 0.0 & 0.0 & 0.0  \\
48 &  0.0 &  0.0 & 0.0 &  0.0 & 0.0 & 0.0 & 0.0 & 0.0 \\
49  & 0.0 & 0.0 & 0.0 & 0.0 & 0.0 & 0.0 & 0.0 &  0.0  \\
50 & 0.0 & 0.0 & 0.0 & 0.0 & 0.0 & 0.0 & 0.0 &  0.0  \\
51  & 0.0 &  0.0 & 0.0 & 0.0 &  0.0 & 0.0 &  0.0 & 0.0  \\
52 & 0.0 & 0.0 & 0.0 & 0.0 &  0.0 &  0.0 & 0.0 & 0.0 \\
53 & 0.0 & 0.0 & 0.0 &  0.0 & 0.0 & 0.0 & 0.0 & 0.0 \\
54 & 0.0 & 0.0 &  0.0 &  0.0 & 0.0 & 0.0 & 0.0 & 0.00006899 \\
55 &  0.01016461 & 0.04471863 & 0.03889865 & 0.09605255 &  0.0 &  0.0 & 0.0 &  0.00002399 \\
56  & 0.0 & 0.0 & 0.0 & 0.0 & 0.0 &  0.0 & 0.0 & 0.0 \\
57 &  0.0 &  0.0 &  0.0 & 0.0 & 0.0 &  0.00000199 & 0.00020995 &  0.01515673 \\
58 & 0.0 &  0.0 & 0.0 & 0.0 & 0.0 &  0.0 & 0.00022594 &  0.00032889 \\
59 & 0.0 & 0.0 & 0.0 &  0.0 &  0.0 & 0.0 & 0.0 &  0.0 \\
60 & 0.0 & 0.0 & 0.0 & 0.0 & 0.0 &  0.0 & 0.0 & 0.0 \\
61  &  0.0 & 0.0 & 0.0 & 0.0 &  - &  - & - & - \\
62 &  0.0 & 0.0 & 0.0 & 0.0 &  0.0 &  0.0 &  0.00010698 & 0.0342186 \\
63 &  0.0 & 0.0 & 0.0 & 0.0 &  0.0 & 0.0 & 0.0 & 0.0 \\
\hline
\end{tabular}
\end{table}

\newpage

\newpage

\newpage

\newpage

\newpage

\end{document}